\documentclass[10pt,journal,compsoc]{IEEEtran}
\ifCLASSOPTIONcompsoc
  \usepackage[nocompress]{cite}
  \usepackage{cite}
\usepackage{amsmath,amssymb,amsfonts}
\usepackage{algorithmic}
\usepackage{graphicx}
\usepackage{textcomp}
\usepackage{xcolor}
\def\BibTeX{{\rm B\kern-.05em{\sc i\kern-.025em b}\kern-.08em
    T\kern-.1667em\lower.7ex\hbox{E}\kern-.125emX}}
\usepackage{diagbox} % 加载宏包
\usepackage{amssymb}
\usepackage{amsfonts}            %数学字体
\usepackage{mathrsfs}            %数学花体
\usepackage{xcolor}
\usepackage{epstopdf}
\usepackage{algorithm}
\usepackage{multirow}
\usepackage{booktabs}
\usepackage{subfigure}
\usepackage{amsthm}

\newtheorem{theorem}{Theorem}  [section]
\def\mykern{\kern-\fboxsep\kern-\fboxrule}
\else
  \usepackage{cite}
\fi
\ifCLASSINFOpdf
\else
\fi
\begin{document}
\title{Hopping-Proof and Fee-Free Pooled Mining in Blockchain}
\author{Hongwei~Shi,
        Shengling~Wang*,~\IEEEmembership{Member,~IEEE,}
        Qin~Hu,
        Xiuzhen~Cheng,~\IEEEmembership{Fellow,~IEEE,}
        Junshan~Zhang,~\IEEEmembership{Fellow,~IEEE,}
        Jiguo~Yu,~\IEEEmembership{Senior Member,~IEEE}
\IEEEcompsocitemizethanks{\IEEEcompsocthanksitem Hongwei Shi and Shengling Wang (Corresponding author) are with the School of Artificial
Intelligence, Beijing Normal University, Beijing, China. \protect\\
E-mail: hongweishi@mail.bnu.edu.cn and wangshengling@bnu.edu.cn
\IEEEcompsocthanksitem Qin Hu is with the Department of Computer and Information Science, Indiana University - Purdue University Indianapolis, IN, USA.
\protect\\
E-mail:qinhu@iu.edu
\IEEEcompsocthanksitem Xiuzhen Cheng is with the Department of Computer Science, The George
Washington University, Washington DC, USA.
\protect\\
E-mail: cheng@gwu.edu
\IEEEcompsocthanksitem Junshan Zhang is with the School of Electrical, Computer and Energy Engineering, Arizona State University, Tempe, AZ 85287 USA.
\protect\\
E-mail: junshan.zhang@asu.edu
\IEEEcompsocthanksitem Jiguo Yu is with the School of Computer Science and Technology, Qilu University of
Technology (Shandong Academy of Sciences), Jinan 250353, China, with the
Shandong Computer Science Center (National Supercomputer Center in Jinan),
Jinan 250014, China, and also with the School of Information Science and
Engineering, Qufu Normal University, Rizhao 276826, China.
\protect\\
E-mail:jiguoyu@sina.com
}
}
\markboth{}%
{Shell \MakeLowercase{\textit{et al.}}: Bare Demo of IEEEtran.cls for Computer Society Journals}
\IEEEtitleabstractindextext{%
\begin{abstract}
The pool-hopping attack  casts down the expected profits of both the mining pool and honest miners in Blockchain.
The mainstream countermeasures,  namely PPS (pay-per-share) and
PPLNS (pay-per-last-N-share), can hedge pool hopping, but  pose a risk to the pool as well as the cost to miners. In this study, we apply the  zero-determinant (ZD) theory  to design a novel pooled mining which offers an incentive mechanism for motivating  non-memorial and memorial evolutionary miners not to switch in pools strategically. In short, our hopping-proof pooled mining has three unique features: 1) {\it fee-free}. No fee is charged if the miner does not hop. 2) {\it wide applicability}. It can be employed in both prepaid  and postpaid  mechanisms.  3) {\it fairness}. Even the
 pool can dominate the game
with any miner, he has to cooperate when the miner does not hop among pools. The fairness of our scheme makes it have
long-term sustainability. To the best of our knowledge, we are the first to propose a hopping-proof pooled mining with the above three natures simultaneously.
Both theoretical and experimental analyses demonstrate the effectiveness of our scheme.
\end{abstract}
}
\maketitle
\IEEEdisplaynontitleabstractindextext
\IEEEpeerreviewmaketitle
\IEEEraisesectionheading{\section{Introduction}\label{intro}}
\IEEEPARstart{B}{lockchain} is the underlying fabric of mainstream cryptocurrency systems such as Bitcoin \cite{2008} and Ethereum \cite{ethereum}. These cryptocurrencies have obtained a phenomenal success, recognized as the {\it wave of future}\cite{stubborn} with a total market capitalization around 179.6B  dollars at present. To realize a distributed and trustable consensus, a data structure
called blockchain\footnote{In this paper, we use Blockchain to denote the technology while blockchain to indicate a chain of blocks.} is introduced in Blockchain system. It is a public ledger including
a sequence of chained {\it blocks}, with each recording a set of
digital transactions. Since anyone can participate in creating and verifying blocks,  Blockchain system  is open,  leading to its  vulnerability.

To deter attacks incurred by  the openness of Blockchain system which  essentially  originates from its decentralized nature, a Proof-of-Work (PoW)\cite{2008} mechanism is employed, which
allows network participants, i.e.,  miners,  can approve new
transactions  only after mining a block successfully, implying that  they need to solve cryptographic puzzles in the form of a hash
computation with success. PoW undoubtedly increases the cost of malicious behavior,  making many security attacks such as Sybil attack financially unaffordable.
This is because 1)
mining is actually a race where only the winner who solves the
PoW task first can verify digital transactions, which
needs a sufficient amount of computational power; 2) solving
cryptographic puzzles is a probabilistic process, implying that
no one would win the race with certainty even though it is
computationally powerful.

In return for mining blocks successfully, miners are rewarded in
proportion to the computational powers they invested. However,   due to significant computational
resources needed and probabilistic factors involved in the mining process, a solo miner has low expected revenue as well as volatility in the reward. For example,  Bitcoin system now  sets the difficulty of mining such that one block is generated every 10 minutes.  Hence, a solo miner  often has to wait
687 days in expectation to mine a  block \cite{bitcoinminingpool}.

To tackle the above issue,  solo miners join coalitions in the form of {\it mining pools}, gathering their computational powers to seek the solution of PoW puzzles and sharing the rewards proportionally to their
contributions.
This undoubtedly increases the chance of solving cryptographic puzzles successfully and makes the mining process more predictable. Hence, pooled mining can benefit miners from  high payoffs and low variance in rewards. At present, nearly 80\% of the computing power in Bitcoin and 60\% of that in Ethereum belong to less than 8 and 3 mining pools, respectively.

The dominant position of pooled mining leads it to become a valuable target to be
attacked. Many pools have an open trait, allowing any miner to join them
through public Internet interfaces\cite{miner}, which makes matters worse. Such a nature of openness makes pooled mining susceptible to attacks. There are mainly three kinds of
security attacks in pooled mining:  the selfish mining attack \cite{optimal,selfish1,selfish2},
the block withholding  attack  \cite{atrans,miner,hu2019} and the pool-hopping
attack  \cite{analysis}. The first two attacks can be well solved through the state-of-the-art approaches  \cite{selfish1,selfish2,splitting,atrans,hu2019}, and hence, we focus on the last one.

The pool-hopping attack was first proposed by Rosenfeld\cite{analysis}, in which the malicious miners strategically switch among the pools to obtain a higher payoff. This attack is cost-efficient and straightforward because of no more extra operations (e.g., keeping the block secret, dropping full proof of work or forking) needed. Studies have proved that the miner has no incentive to stay in a pool without pool hopping or redistributing the computing power\cite{bitcoinminingpool,splitting,atrans}. This kind of greedy and   opportunistic manner definitely casts down the mining power of a pool, resulting in its declined expected revenue. In addition, the pool-hopping attack also jeopardizes the interests of honest miners, who join in a pool continuously without switching to other pools. According to \cite{analysis}, the honest miners in the attacked pool will receive 43\% less payoff in the worst-case theoretically, which is unfair for them.

However, little research has studied the pool-hopping attack. PPS and
PPLNS \cite{analysis} are pioneer countermeasures. Considering that the unbalanced distribution of reward over time makes room for miners' strategic hopping, the key idea of  PPS and  PPLNS is reducing the variance of reward in time series. Typically, in a PPS pool, a miner will be rewarded as long as she\footnote{In this paper, we denote the pool as ``he" and the miner as ``she" for easy differentiation.} submits a share (her contribution) to the pool, regardless of whether a block is mined successfully or not. PPLNS, one of the most prevailing reward mechanisms\cite{diver}, drops the concept of ``round", focusing on  $N$ shares submitted to the pool recently and distributing  rewards according to the shares in proportion.

Essentially, the difference between PPS and PPLNS lies in that the former is driven by events while the latter is triggered by time. In detail, PPS rewards a miner once the event of receiving her share happens; PPLNS evaluates whether a miner should be awarded when the paying time arrives. The common feature of PPS and PPLNS is that they pay miners  proportionally to their contribution, regardless of whether a block is mined successfully or not. Due to the uncertainty of mining results, the pool takes the full risk when no block is mined. Therefore,  both PPS and PPLNS charge miners some fees to  alleviate such a risk, which is critical to both of the pool and its members.
The higher the fee, the higher the cost of the miner joining in the pool, and the smaller the motivation to mine and vice versa.

In a nutshell, the mainstream countermeasures to the pool-hopping attack, namely PPS and PPLNS, pose a risk to the pool as well as the cost to miners. Therefore, we propose a hopping-proof pooled mining  with free fee in this paper, which can hedge pool hopping without any fee charged if the miner does not switch in pools strategically. The proposed pooled mining  strategy has a wide scope of application since it can be employed in both prepaid  and postpaid  mechanisms. The former rewards once share is submitted,  no matter whether there is a success mining or not;  the latter awards only when the full cryptographic puzzle is solved.

It is challenging to realize the
 hopping-proof pooled mining without fee charged. The reasons behind the fact are: a) the strategic transferring among different pools is the instinctive demand of a miner. Especially when no fee is charged, costless hopping easily arouses miners to switch among pools; b) in the postpaid mode, mining risk is completely transferred from the pool to miners. In this situation,  it is non-trivial to motivate miners to still work without hopping.

To tackle these challenges, we take advantage of the zero-determinant (ZD) theory to design an incentive mechanism for pooled mining, where cooperation (i.e., mining without hopping) is the dominant strategy of a rational miner in all situations. The ZD theory was first developed in \cite{PNAZ} by Press and Dyson, in which the player who adopts the ZD strategy (i.e., the ZD adopter) can unilaterally set its adversary's utility no matter what strategy the adversary takes. The power of the ZD strategy endows the pool to dominate the game with any miner, rewarding her cooperation and punishing the defection, to lure the cooperation of the miner.

The main contributions of this paper can be summarized as follows:
\begin{itemize}
  \item The interaction between the pool and any miner is formulated as an iterated prisoner's dilemma (IPD) game and the corresponding conditions are also identified. The generality of our model empowers the proposed pooled mining to have a wide scope of application, implying that it is suitable to  both prepaid  and postpaid  mechanisms. When applied in a postpaid mechanism, the proposed pooled mining can incentivize miners to work without hopping while keeping the pool away from the risk of no block mined successfully.
  \item  We investigate in detail whether the pool can be a ZD adopter and how
he plays the ZD strategy. We draw a conclusion that the pool can unilaterally control the miner's payoff rather than his own one. The specific expected payoff of a miner that the pool can set is characterized.
  \item An incentive mechanism based on the ZD theory is proposed for motivating the non-memorial and memorial evolutionary miners to work without hopping. Specifically, the proposed mechanism empowers the pool to encourage the miner to behave cooperatively by increasing her short-term payoff without any additional payment in the long run.
  \item Both theoretical and experimental analyses demonstrate the effectiveness of the proposed incentive mechanism. More importantly, we find the proposed pooled mining is fair, implying that even the pool can dominate the game with any miner, he has to cooperate when the miner works collaboratively. The fairness of our scheme makes it have long-term sustainability.
  \end{itemize}

The rest of the paper is organized as follows. Section \ref{sec:sec3} describes the formulation of our problem. The ZD strategy for the pool in an iterated prisoner's game is deduced in Section \ref{sec:sec4}. Based on which, we propose an incentive mechanism in light of the ZD theory in Section \ref{sec:sec 5}. We evaluate the mechanism both theoretically in Section \ref{sec:sec6} and experimentally in Section \ref{sec:evalu}. The related literatures are listed in Section \ref{sec:sec2}. Section \ref{sec:sec_conclude} concludes our paper finally.
\section{Game Formulation}\label{sec:sec3}
In this section, we introduce our game model to formulate the interaction between the pool and the miner. Generally, we define the strategy space of each player as a dichotomous space, namely cooperation ($c$) and defection ($d$). In the PoW mining scenario, the pool is considered as a cooperator if he decides to pay the highest payoff to the miner; otherwise, he is regarded as a defector. On the other hand, the miner can devote herself wholeheartedly to the current pool by providing her total computational power to the pool without hopping, defined as cooperation, or contribute herself halfheartedly through offering partial computing ability or switching to other pools strategically, denoted by defection. We denote the actions of the pool and the miner as $x,y\in\{c,d\}$, respectively. Therefore, there are four possibilities of states in each round between the pool and the miner, i.e., $XY=(cc,cd,dc,dd)$, where $X$ and $Y$ denote the state of the pool and that of the miner, respectively. It is worth to note that the terminal of a mining round mentioned in our model can be defined as the time a block is mined successfully or the paying time similar to that in PPLNS. Hence,  the proposed scheme can be applied in  both prepaid  and postpaid  mechanisms.

Each state will correspond to specific payoffs for both players, which can be derived as follows:
\begin{itemize}
\item if both the pool and the miner are collaborative with the pool providing the highest payoff and the miner offering her entire computing power to the current pool, the payoffs of them are represented as $K_p$ and $K_m$, respectively;
\item when the miner defects while the pool cooperates, the miner will get an increase of $\sigma>0$ based on her original payoff $K_m$, while the pool may obtain a decrease of $\pi>0$ on $K_p$;
\item in the case that the defective pool plays against a cooperative miner, the payoff of the pool increases by $\mu>0$, while the miner receives a loss of $\rho>0$;
\item when  both players behave maliciously, the payoffs of the pool and the miner are  $K_p-\pi+\mu$ and $K_m+\sigma-\rho$, respectively.
\end{itemize}

Subsequently, the payoff vectors of the pool, denoted as $\mathbf S_p=(S_p^{xy})$, and the miner, denoted as $\mathbf S_m=(S_m^{xy})$, $x,y \in \{c, d\}$, can be presented as follows $$\mathbf S_p=(S_p^{cc},S_p^{cd},S_p^{dc},S_p^{dd})=(K_p,K_p-\pi,K_p+\mu,K_p-\pi+\mu),$$
$$\mathbf S_m=(S_m^{cc},S_m^{cd},S_m^{dc},S_m^{dd})=(K_m,K_m+\sigma,K_m-\rho,K_m+\sigma-\rho),$$
which are also shown in Table \ref{tb:payoff}.
%payoff矩阵
\begin{table}[!htbp]
\caption{Payoff matrix of the pool and the miner}
\centering
\begin{tabular}{|c|c|c|}
\hline
\diagbox{Pool}{Miner}&Cooperation&Defection\\ % 添加斜线表头
\hline
Cooperation&$K_p,K_m$&$K_p-\pi,K_m+\sigma$\\
\hline
Defection&$K_p+\mu,K_m-\rho$&$K_p-\pi+\mu,K_m+\sigma-\rho$\\
\hline
\end{tabular}
\label{tb:payoff}
\end{table}

Next, some insightful theorems are introduced to characterize the game in the following.

\begin{theorem}\label{2.1}
{\it If $\pi>\mu, \rho>\sigma, \mu<\rho, \sigma<\pi$, a prisoner's dilemma (PD) game can be modeled to depict the confrontation between the pool and the miner. }
\end{theorem}
\begin{IEEEproof}
To become a PD game, two fundamental conditions should be satisfied. In detail, 1) the stable state occurs when both players defect, i.e., $XY=dd$ is the Nash equilibrium; 2) mutual cooperation is the best outcome with respect to the social welfare, which means $XY=cc$ outperforms other states from an overall perspective.

The game between the pool and the miner satisfies the first condition. To be specific, if the miner is friendly, the pool will get a lower payoff as $K_p$ when he cooperates than his payoff of $K_p+\mu$ when he defects; besides, if the pool challenges with a malicious miner, the payoff when he defects, i.e., $ K_p-\pi+\mu $, is also larger than that of his cooperation, i.e., $K_p-\pi$. Thus, as a rational decision maker, the pool will always choose to defect rather than cooperation when facing an adversary with uncertain actions. With similar analysis, we can find the only feasible option for a rational miner is also to behave viciously. Accordingly, both the pool and the miner will select defection as the stable state. Therefore, the Nash equilibrium of this game comes to be $XY=dd.$

In order to investigate the second condition clearly, we denote the social welfare in each state as $W_{cc}, ~W_{cd}, ~W_{dc}$ and $W_{dd}$. Thus, we have $W_{cc}=K_p+K_m$,
$W_{cd}=K_p+K_m+\sigma-\pi$, $W_{dc}=K_p+K_m-\rho+\mu$, and $W_{dd}=K_p+K_m+\sigma+\mu-\rho-\pi$. Then the second condition is satisfied when $W_{cc}>W_{cd}, W_{cc}>W_{dc}, W_{cc}>W_{dd}$ hold.
It is obvious that when $\pi>\mu, \rho>\sigma, \mu<\rho, \sigma<\pi$, the above inequalities can be satisfied.
Based on the analyses above, as self-regarding players, the pool and the miner will choose malicious behavior to maximize their payoffs, leading to mutual defection as the stable state in the game consequently. However, the most favorable outcome of the confrontation turns out to be mutual cooperation. Therefore, a PD game is formed when $\pi>\mu, \rho>\sigma, \mu<\rho, \sigma<\pi$.
\end{IEEEproof}

Notably, the miner may stay in the current pool for a long time without hopping to others. Hence, in this case, the PD game mentioned above can become an iterated one if some conditions are satisfied, which are summarized in the following theorem.

\begin{theorem}\label{2.2}
{\it If $\pi>\mu, \rho>\sigma, \mu<\rho, \sigma<\pi$, the confrontation between the pool and the miner can be modeled as an iterated prisoner's dilemma (IPD) game. }
\end{theorem}
\begin{IEEEproof}
A PD game becomes an iterated one when the payoff of any player's persistence on cooperation is larger than hopping between cooperation and defection.
%insisting in state $XY=cc$ excels those in states $XY=cd$ and $XY=dc$.
In other words, the inequalities below should hold
\begin{align} \label{eq:IPD}
\left\{
\begin{aligned}
    2K_p>K_p+\mu+K_p-\pi,\\
    2K_m>K_m+\sigma+K_m-\rho.\\
\end{aligned}
\right.
\end{align}
Hence, when $\pi>\mu, \rho>\sigma, \mu<\rho$ and $\sigma<\pi$, the game between the pool and the miner can be modeled as an IPD one.
\end{IEEEproof}

In light of the above analyses, we can find that the miner and the pool may be trapped into the iterated prisoner's dilemma, where the Nash equilibrium is far away from mutual cooperation, leading to low efficiency and distrust for  Blockchain system in the long run. To tackle this problem, we employ the powerful ZD strategy to drive the players to cooperate so as to reach the win-win situation. As introduced in Section \ref{intro},  the ZD adopter can unilaterally set its adversary's payoff no matter what strategy the adversary takes.

Aware of such an effective strategy, the pool is attracted to use the ZD strategy to resist a hopping miner. In this case, however, we are facing the following problems: {\it is the pool capable of being a ZD adopter? if yes, how does the ZD strategy work? }To address these questions, we conduct the following analyses.
\section{ZD Strategy for the Pool}\label{sec:sec4}
In this section, we examine whether the pool can play the ZD strategy, and if yes, how to achieve that. %according to \cite{PNAZ}.
Firstly, a Markov game is established between the pool and the miner.  As mentioned in Section \ref{sec:sec3}, there are four possible game results, i.e., $XY=(cc,cd,dc,dd)$, in each round. We define the pool's mixed strategy as $\mathbf p=(p_1,p_2,p_3,p_4)$, where $p_1$ represents the probability of choosing cooperation in this round based on the previous outcome $cc$. %\footnote{$X_{-1}Y_{-1}$ denotes the game state in the last round.}
Similarly, when the previous outcome is $cd$, $dc$ or $dd$, the probability of the pool to cooperate in this round is $p_2$, $p_3$ or $p_4$. Accordingly, the probability of the pool being defective in each round is $(1-p_1,1-p_2,1-p_3,1-p_4)$ corresponding to different game results in last round. Comparably, in the cases that the miner chooses to cooperate when $cc, cd,dc$ or $dd$ happens previously, her strategy can be denoted as $\mathbf q=(q_1,q_2,q_3,q_4)$, while the probability of defecting is $(1-q_1,1-q_2,1-q_3,1-q_4).$

With the above-defined strategies of the pool and the miner, the Markov matrix  in each round can be derived as follow,%$\mathbf A$, that is,
$$
\mathbf A=\left[
  \begin{matrix}
    p_1q_1&p_1(1-q_1)&(1-p_1)q_1&(1-p_1)(1-q_1)\\
    p_2q_2&p_2(1-q_2)&(1-p_2)q_2&(1-p_2)(1-q_2)\\
    p_3q_3&p_3(1-q_3)&(1-p_3)q_3&(1-p_3)(1-q_3)\\
    p_4q_4&p_4(1-q_4)&(1-p_4)q_4&(1-p_4)(1-q_4)
  \end{matrix}
  \right],
$$
where each element denotes the probability of state transition. For example, if the previous outcome is $cc$, combining the cooperation probabilities of the pool and the miner, i.e., $p_1$ and $q_1$, the probability of $XY=cc$ in this round is $p_1q_1$, so do other elements in $\mathbf A$.

Denote $\mathbf v$ as the stationary vector of matrix $\mathbf A$, then $\mathbf v^T\mathbf A=\mathbf v^T$ and $\mathbf v^T\mathbf M=\mathbf 0$, where $\mathbf M=\mathbf A-\mathbf I$ ($\mathbf I$ is the identity matrix). According to the Cramer's rule, the equation $Adj(\mathbf M)\mathbf M=det(\mathbf M)\mathbf I= \mathbf 0$ holds, where $Adj(\mathbf M)$ and $det(\mathbf M)$ represent the adjugate matrix and the determinant of $\mathbf M$. Subsequently, the equation above indicates that every row of $Adj(\mathbf M)$ is in proportion to $\mathbf v$\cite{PNAZ}. Thus, if the dot product of $\mathbf v$ with any vector $\mathbf f=(f_1,f_2,f_3,f_4)^T$ is conducted, the determinate remains unchanged with some elementary column transformation, such as adding the first column to the second and the third columns. Thus, we have,
$$
\mathbf v \cdot \mathbf f = D(\mathbf p, \mathbf q,\mathbf f) =det \left[
  \begin{matrix}
    p_1q_1-1&p_1-1&q_1-1&f_1\\
    p_2q_2&p_2-1&q_2&f_2\\
    p_3q_3&p_3&q_3-1&f_3\\
    p_4q_4&p_4&q_4&f_4
  \end{matrix}
  \right].
$$

It is evident that the second column of the above determinant is only related to the pool's strategy. Based on this, the expected payoffs of the pool ($ S_p$) and the miner ($S_m$) can be derived as
\begin{equation}\label{eq:payoff_original}
S_p=\frac{\mathbf {v} \cdot \mathbf S_p}{\mathbf {v} \cdot \mathbf 1}=\frac{D(\mathbf p,\mathbf q,\mathbf S_p)}{D(\mathbf p,\mathbf q,\mathbf 1)},\notag
\end{equation}

\begin{equation}\label{eq:payoff original-Sy}
S_m=\frac{\mathbf {v} \cdot \mathbf S_m}{\mathbf {v} \cdot \mathbf 1}=\frac{D(\mathbf p,\mathbf q,\mathbf S_m)}{D(\mathbf p,\mathbf q,\mathbf 1)}.
\end{equation}

Hence, the linear relationship between the pool and the miner's expected payoffs holds as follows
\begin{equation}\label{eq:zd1}
\alpha S_p + \beta S_m + \gamma =\frac{D(\mathbf p,\mathbf q,\mathbf \alpha \mathbf S_p + \beta \mathbf S_m + \gamma\mathbf 1)}{D(\mathbf p,\mathbf q,\mathbf 1)},
\end{equation}
where $\alpha,~\beta,~\gamma$ are coefficients.

Therefore, if the pool sets his strategy the same as $\mathbf \alpha \mathbf S_p + \beta \mathbf S_m + \gamma\mathbf 1$, the determinant in the numerator equals 0, because there exists two identical columns. In this case, $\alpha S_p + \beta S_m + \gamma=0$, implying that a linear relation is established between the expected payoffs $S_p$ and $S_m$, where the corresponding strategy is therefore called {\it Zero-Determinant Strategy}, denoted as $\hat{\mathbf p}$  below.

Specifically, when the pool sets $\hat{\mathbf p}=\beta \mathbf S_m + \gamma\mathbf 1$ (i.e., $\alpha=0$), the pool can control the miner's expected payoff independently as $S_m=-\frac{\gamma}{\beta}$; while when he exerts his strategy as $\hat{\mathbf p}=\alpha \mathbf S_p + \gamma\mathbf 1$ by setting $\beta=0$, he can set his own expected payoff at $S_p=-\frac{\gamma}{\alpha}$. %, the pool can make his strategy as $\hat{\mathbf p}=\alpha \mathbf S_p + \gamma\mathbf 1$, by setting $\beta=0$. Accordingly, the pool can utilize the powerful ZD strategy based on his strategy $\hat{\mathbf p}$.
The following theorem demonstrates the effectiveness of the ZD strategy adopted by the pool.
%\subsection{Analysis of the pool being a ZD player}

\begin{theorem}\label{3.1}
{\it The pool can unilaterally control the miner's expected payoff as $S_m=\frac{(1-p_1)S_m^{dd}+p_4S_m^{cc}}{1-p_1+p_4}$, while he is not able to set his own expected payoff independently.}
\end{theorem}
\begin{IEEEproof}
Firstly, if the pool wants to control his adversary's  expected payoff as $S_m=-\frac{\gamma}{\beta}$ by setting $\alpha=0$, the specific ZD strategy of the pool should satisfy $\hat{\mathbf p}=\beta \mathbf{S_m}+\gamma\mathbf 1$, %, that is:
%\begin{align} \label{eq:pool_con_miner}
%\left\{
%\begin{aligned}
%    p_1-1&=\beta S_m^{cc}+\gamma, \\
%    p_2-1&=\beta S_m^{cd}+\gamma, \\
%    p_3&=\beta S_m^{dc}+\gamma, \\
%    p_4&=\beta S_m^{dd}+\gamma.
%\end{aligned}
%\right.
%\end{align}
%Based on equations \eqref{eq:pool_con_miner}, we have $$\beta=\frac{p_1-p_4-1}{S_m^1-S_m^4},\gamma=\frac{(1-p_1)S_m^4+p_4S_m^1}{S_m^1-S_m^4}.$$
according to which, we can deduce $p_2$ and $p_3$ with respect to $p_1$ and $p_4$,
\begin{align}\label{eq:pool_con_miner_p_23}
\left\{
\begin{aligned}
    p_2=\frac{p_1(S_m^{cd}-S_m^{dd})-(1+p_4)(S_m^{cd}-S_m^{cc})}{S_m^{cc}-S_m^{dd}},\\
    p_3=\frac{(1-p_1)(S_m^{dd}-S_m^{dc})+p_4(S_m^{cc}-S_m^{dc})}{S_m^{cc}-S_m^{dd}}.\\
\end{aligned}
\right.
\end{align}

It is evident that $p_2$ and $p_3$ are meaningful as they belong to $[0,1].$ Therefore, it is clear that being a ZD player, the pool can set the miner's expected payoff unilaterally. And the miner's expected payoff comes to be \begin{equation}\label{sm}
  S_m=-\frac{\gamma}{\beta}=\frac{(1-p_1)S_m^{dd}+p_4S_m^{cc}}{1-p_1+p_4}.
\end{equation}

As \eqref{sm} consisting of a weighted average of $S_m^{cc}$ and $S_m^{dd}$ with weights $p_4$ and $1-p_1$, we can conclude that the expected payoff of the miner can be set in the range of $[S_m^{dd}, S_m^{cc}]$ by the pool's ZD strategy.

Secondly, when it comes to the case that the pool sets his own expected payoff, the ZD adopter's strategy should meet  $\hat{\mathbf p}=\alpha \mathbf{S_p}+\gamma\mathbf 1$ ($\beta =0$). %that is to say, the following equation set should be satisfied:
%\begin{align} \label{eq:pool_con_pool}
%\left\{
%\begin{aligned}
%    p_1-1&=\alpha S_p^{cc}+\gamma, \\
%    p_2-1&=\alpha S_p^{cd}+\gamma, \\
%    p_3&=\alpha S_p^{dc}+\gamma, \\
%    p_4&=\alpha S_p^{dd}+\gamma.
%\end{aligned}
%\right.
%\end{align}
Using $p_1$ and $p_4$ to represent $\alpha$ and $\gamma$, we have
\begin{align}\label{eq:a+r}
\left\{
\begin{aligned}
    \alpha&=\frac{p_1-p_4-1}{S_p^{cc}-S_p^{dd}},\\
    \gamma&=\frac{(1-p_1)S_p^{dd}+p_4S_p^{cc}}{S_p^{cc}-S_p^{dd}}.
\end{aligned}
\right.
\end{align}

And we can use $p_1$ and $p_4$ to describe $p_2$ and $p_3$ as
\begin{align}\label{eq:pool_con_pool2}
\left\{
\begin{aligned}
    p_2&=\frac{(1+p_4)(S_p^{cc}-S_p^{cd})-p_1(S_p^{dd}-S_p^{cd})}{S_p^{cc}-S_p^{dd}},\\
    p_3&=\frac{-(1-p_1)(S_p^{dc}-S_p^{dd})-p_4(S_p^{dc}-S_p^{cc})}{S_p^{cc}-S_p^{dd}},\\
\end{aligned}
\right.
\end{align}
which indicates $p_2\geq1$ and $p_3\leq0$. % according to equation \eqref{eq:pool_con_pool2}.
Under this condition, the pool's strategy is feasible in only one case, i.e., $\hat{\mathbf p}=(1,1,0,0)$, resulting in $\alpha=0$ and $\gamma=0$ according to \eqref{eq:a+r}. Thus, as a ZD player, the pool cannot control his payoff.
\end{IEEEproof}
\section{Incentive Mechanism based on the ZD Strategy}\label{sec:sec 5}
In this section, we propose a ZD-based incentive mechanism for the pooled mining to hinder pool-hopping attacks. Theorem \ref{3.1} reveals the capability of the pool as a ZD player to  set the miner's expected payoff unilaterally.  However, whether the pool can take advantage of such a capability to regulate the miner depends on  her strategy. If the miner's strategy is irrelevant to her payoff, such as   all-cooperation (ALLC, $\mathbf q=(1,1,1,1)$), all-defection (ALLD, $\mathbf q=(0,0,0,0)$), tit-for-tat (TFT, $\mathbf q=(1,1,0,0)$), the pool cannot employ the ZD strategy to motivate the cooperative behavior of the miner. Hence, the proposed ZD-based incentive mechanism is suitable for the case that the  strategy is laid down by the miner  in light of her payoff. Win-stay-lose-shift (WSLS, $\mathbf q=(1,0,0,1)$) and evolutionary strategies are typical  payoff-driven examples.

%  To begin with, we analyze the situations where the miner adopts four classical strategies, including all-cooperation (ALLC, $\mathbf q=(1,1,1,1)$), all-defection (ALLD, $\mathbf q=(0,0,0,0)$), tit-for-tat (TFT, $\mathbf q=(1,1,0,0)$) and win-stay-lose-shift (WSLS, $\mathbf q=(1,0,0,1)$). %In the following, we introduce four strategies in detail.

%In the cases where the miner adopts ALLC and ALLD, the only feasible strategy for the miner is to cooperate and defect irrespective of what strategy the pool would employ. Hence, in this paper, incentive mechanisms are not considered for such fixed and inveterate strategies due to their steady characteristics. Besides, when the pool cooperates first, it is conceivable that the miner who employs TFT strategy will choose collaborative behavior in the following, resulting in mutual cooperation in the end. Thus, it is easy for the pool to encourage cooperative action of a TFT miner through cooperating firstly.

%Additionally, %the win-stay-lose-shift strategy shares similarities with the evolutionary strategy.(No introduction of evolutionary strategy here.) To be specific, WSLS is defined as:
A WSLS player will keep the same strategy as the previous round in which the outcome  is good,  that is so called ``win-stay''. Otherwise, it will adopt the strategy opposite to the one in the previous round, which is therefore named as ``lose-shift''.   Hence, WSLS can be regarded as a particular case of the evolutionary strategy. In this work, we take the evolutionary strategy as the representative for further analysis, which can be categorized into two kinds: {\it non-memorial} and {\it memorial}. We introduce them in detail as follows.

\subsection{Evolutionary strategies}
%Firstly, we introduce the non-memorial and memorial evolutionary strategies for the miner. In light of the evolutionary nature of the miner, the pool is capable of developing incentive mechanisms based on the ZD theory to encourage cooperative behavior.
The non-memorial evolutionary (\textbf{E}) strategy is featured by the fact that an \textbf{E} player may develop the strategy only based on its expected payoff. Specifically, as a rational player, if the cooperative behavior brings about a higher payoff than the defective one, the \textbf{E} player will choose to collaborate and vice versa. A typical non-memorial evolutionary strategy can be formulated as follow \cite{young},
\begin{align} \label{eq:10}
\begin{aligned}
    q^t(c|\mathbf p)=\frac{e^{\epsilon[E_m^t(c|\mathbf p )-E_m^t(d|\mathbf p)] }}{1+e^{\epsilon[E_m^t(c|\mathbf p )-E_m^t(d|\mathbf p)] }}, \end{aligned}
\end{align}
where $q^t(c|\mathbf p)$ denotes the non-memorial \textbf{E} player's cooperation probability in round $t$ based on the pool's strategy $\mathbf p$ and $\epsilon>0$ is a scaling parameter. Besides,  $E_m^t(c|\mathbf p )$ and $E_m^t(d|\mathbf p )$ represent the expected payoffs of the miner who acts cooperatively and defectively. %when the pool plays strategy $\mathbf p$ in round $t$.
%In detail, $E_m^t(c|\mathbf p )$ ($E_m^t(d|\mathbf p )$) equals to the total payoffs when the miner cooperated (defected) previously divided by the number of rounds she collaborated (defected).
%; similarly, $E_m(d|\mathbf p )$ can be derived as the payoff when the miner defects divided by the number of defective rounds, i.e.,
%\begin{equation}
%\begin{aligned}
%E_m(c|\mathbf p )=\Sigma_{i=1}^M\frac{u_m(c|\mathbf p^i )}{N_c},\\
%E_m(d|\mathbf p )=\Sigma_{i=1}^M\frac{u_m(d|\mathbf p^i )}{N_d}.
%\end{aligned}
%\end{equation}

%In the above equations,  $u_m(c|\mathbf p^i)$ and $u_m(d|\mathbf p^i)$ indicate the miner's payoffs when she cooperates and defects in each round $i$, which can be controlled by the pool via employing ZD strategy. Besides, $N_c$ and $N_d$ represent the number of rounds the miner collaborates and defects.
%From \eqref{eq:28}, it is clear that if the miner obtains more utility as a cooperative player, the probability of cooperating will be increased. That is to say, the miner is more likely to devote her computing power entirely to the pool. Therefore, as a ZD player, the pool may reward the cooperation of a miner with a higher payoff while punishing her defection with the lower one. The above reward-punishment mechanism facilitates to incentivize cooperation to a great extent.
%\subsection{Memorial evolutionary strategy}
Different from the non-memorial evolutionary strategy, the memorial evolutionary strategy is associated with not only the expected payoff but also its strategy in the previous round, which we call it {\it memory}. That is to say, informed of the previous strategy and the expected payoff, the memorial \textbf{E} player may adjust its strategy more rationally.

Inspired by \cite{evolutionary}, we present the memorial evolutionary strategy as following: if the cooperation probabilities of the pool and the miner are denoted as $p^t$ and $q^t$ in round $t$, then the miner's cooperation probability $q^{t+1}$ in the next round evolves as
\begin{align} \label{eq:memo}
\begin{aligned}
    q^{t+1}=q^t\cdot \frac{W_c^t}{E_m^t}, \end{aligned}
\end{align}
where $W_c^t$ indicates the expected payoff of the miner when she cooperates and $E_m^t$ implies the expected payoff of the miner in round $t$. Accordingly, $W_c^t$ and $E_m^t$ can be calculated by
\begin{equation}\label{eq:12}
\begin{aligned}
W_c^t&=p^t\cdot S_m^{cc}+(1-p^t)\cdot S_m^{dc},\\
E_m^t&=q^t\cdot W_c^t+(1-q^t)\cdot W_d^t,
\end{aligned}
\end{equation}
where $W_d^t=p^t\cdot S_m^{cd}+(1-p^t)\cdot S_m^{dd}$ is the miner's expected payoff when she defects.

\subsection{ZD incentive mechanism}
From equations \eqref{eq:10} and \eqref{eq:memo}, it is clear that if the miner obtains more payoff as a cooperative player, her cooperation probability will increase. That is to say, the miner is more likely to devote her computing power entirely to the pool without hopping if such an action brings about a higher payoff. Therefore, as a ZD player, the pool may reward the cooperation of a miner with a higher payoff while punishing her defection with the lower one. Based on this, we propose a  ZD-based incentive mechanism for the pool to coerce the miner's collaborative action, thereby deterring the hopping behavior of the miner, which is detailed in the following.

As shown in Algorithm 1,  in the first round, we offer the reward to each miner $i$ ($i=1,2,...,N$) proportionally to her contribution to the pool. The historical best computing power $B_i$ is recorded as the initial computation power of each miner $i$, namely $m_i^1$ (Lines 1-4). In practice, whether a miner behaves cooperatively or defectively can not be deduced  without any side information, since it is the private information of the miner.  Hence, the pool has to  differentiate a collaborate or defective miner based on the observation of the difference of computational powers between two continuous rounds. This requires the pool  to record the computation power  $m_i^j$ of any miner $i$  at the end of each round $j$ (Line 7),   so that
the pool can obtain the difference  of the devoted computational power of miner $i$ between round $j-1$ and round $j$, i.e.,  $\Delta m_i^j=m_i^j-m_i^{j-1}$ (Line 8). If  $\Delta m_i^j\geq 0$, miner $i$ is considered to be  a cooperative player and vice versa.

When $\Delta m_i^j<0$,  the miner splits her computing power into other pools\footnote{The situation where the miner is unavailable due to some reasons such as lacking of electricity is out of our consideration in this paper.}, implying she is a pool-hopping attacker. Her payoff  is therefore needed to be reduced in order to hinder such an attack. Under this situation, the pool will exert the ZD strategy, setting the attacker's payoff as the minimum one, i.e., $L$ (Lines 9-10). If  $\Delta m_i^j=0$,  the pool  provides the same payoff to the miner as that in  the last round (Lines 11-12). When $\Delta m_i^j>0$,  the pool would update $B_i$ if needed (Lines 14-16). Since this case indicates the miner behaves more cooperatively, the pool will increase her  payoff as  $E_i^j=H*\frac{e^{\zeta\cdot y}}{1+e^{\zeta\cdot y}}$, where $y=(\frac{\Delta m_i^j}{B_i}+1)\cdot E_i^{j-1}$ and $\zeta>0$ represents a scaling parameter (Line 17-18). It is worth to note that the more increment of computational power relative to  $B_i$ is, the higher reward the miner can obtain, which is up to  the maximum payoff that the pool can offer, namely $H$.

\begin{algorithm}[t]\label{al1}
\caption{The ZD-based incentive mechanism}
\begin{algorithmic}[1]\REQUIRE ~~\\ %算法的输入参数：Input
The total number of iterations, $M$;\\
The number of miners, $N$;\\
The initial computation power of miner $i$, $m_i^1$;\\
The minimum and maximum payoffs that the pool can offer, $L$ and $H$;\\
\FOR{$i=1$ to $N$}
    \STATE Calculate the initial reward according to $\frac{m_i^1}{\sum_{i=1}^N m_i^1}\cdot[H-L]+L$
    \STATE $B_i=m_i^1$
\ENDFOR
\FOR{$i=1$ to $N$}
\FOR{$j=2$ to $M$}
    \STATE Update computation power $m_i^j$
    \STATE $\Delta m_i^j=m_i^j-m_i^{j-1}$
    \IF{$\Delta m_i^j<0$}
        \STATE Calculate $\mathbf p^j$ which makes $E_i^j=L$
    \ELSIF{$\Delta m_i^j=0$}
        \STATE  $\mathbf p^j=\mathbf p^{j-1}$ which makes $E_i^j=E_i^{j-1}$
    \ELSIF{$\Delta m_i^j>0$}
     \IF {$B_i<m_i^j$}
            \STATE $B_i=m_i^j$
        \ENDIF
        \STATE $y=(\frac{\Delta m_i^j}{B_i}+1)\cdot E_i^{j-1}$
        \STATE Calculate $\mathbf p^j$ which makes $E_i^j=H*\frac{e^{\zeta\cdot y}}{1+e^{\zeta\cdot y}}$
           \ENDIF
\ENDFOR
\ENDFOR
\end{algorithmic}
\end{algorithm}
\section{Theoretical Analysis}\label{sec:sec6}
In this section, we analyze the proposed incentive mechanism theoretically.

\begin{theorem}\label{5.1}
{\it For any non-memorial evolutionary miner who is motivated by the ZD incentive mechanism, it is conceivable that the miner's cooperation probability will be maximized.}
\end{theorem}
\begin{IEEEproof}
To maximize $q^t(c|\mathbf p)$ according to \eqref{eq:10}, we turn to prove that $E_m^t(c|\mathbf p )-E_m^t(d|\mathbf p)$ rises with the increase of game round $t$ if the miner is a cooperative one. According to Algorithm 1, if any miner $i$ behaves  more cooperatively than the previous round, we have

\begin{equation}\label{eq:14}
\begin{aligned}
E_m^t(c|\mathbf p )-E_m^t(d|\mathbf p)&=H*\frac{e^{\zeta\cdot y}}{1+e^{\zeta\cdot y}}-L\\
&=H*\frac{e^{\zeta\cdot (\frac{\Delta m_i^t}{B_i}+1)\cdot R_i^{t-1}}}{1+e^{\zeta\cdot (\frac{\Delta m_i^t}{B_i}+1)\cdot R_i^{t-1}}}-L.
\end{aligned}
\end{equation}

Since $(\frac{\Delta m_i^t}{B_i}+1)\cdot R_i^{t-1}$ keeps raising because of the miner's collaborative behavior, $\frac{e^{\zeta\cdot (\frac{\Delta m_i^t}{B_i}+1)\cdot R_i^{t-1}}}{1+e^{\zeta\cdot (\frac{\Delta m_i^t}{B_i}+1)\cdot R_i^{t-1}}}$ becomes to one at last, leading $E_m^t(c|\mathbf p )-E_m^t(d|\mathbf p)$ equals to $H-L$ consequently. Hence,  driven by the proposed ZD incentive mechanism,  $q^t(c|\mathbf p)$  can evolve to the maximum.
\end{IEEEproof}
%For a non-memorial evolutionary miner, her cooperation probability $q(c|\mathbf p)$ increases when the difference between $E_m(c|\mathbf p )$ and $E_m(d|\mathbf p )$ raises as shown in \eqref{eq:10}. Not surprisingly, the largest difference between $E_m(c|\mathbf p )$ and $E_m(d|\mathbf p )$ happens when the pool employs the extreme ZD incentive mechanism to award the collaborative miner with the highest payoff while punishing her with the lowest one. Consequently, the miner will set her cooperative possibility increasingly, leading to the collaborative behavior in the end.

%As for the case where the non-memorial evolutionary miner is driven by the proportional ZD incentive mechanism, we can draw the same conclusion since the only difference between these two incentive mechanisms lies in the way how the pool reward a cooperative miner. Although the pool will exert his ZD strategy to set the miner's payoff in proportion to the number of rounds she cooperates, rather than providing the maximal payoff directly, it is evident that the miner is coerced to behave cooperatively with probability 1 as her payoff increases when she cooperates.

%To sum up, the extreme ZD incentive mechanism and the proportional ZD incentive mechanism are capable of motivating a non-memorial evolutionary miner to behave collaboratively with probability one at last.

\begin{theorem}\label{5.2}
{\it For any memorial evolutionary miner who is motivated by the ZD incentive mechanism, her cooperation probability tends to 1 gradually.
}
\end{theorem}
\begin{IEEEproof}
In light of \eqref{eq:memo}, a memorial evolutionary miner can calculate her cooperation probability according to $W_c^t$ and $E_m^t$, which can be deduced by \eqref{eq:12}. In practice, we use the cooperative frequencies $f_p^t$ and $f_m^t$ to approximate $p^t$ and $q^t$. Specifically, $f_p^t$ indicates the number of rounds the pool cooperates divided by the total number of rounds, while $f_m^t$ denotes that of a miner.

Based on the ZD incentive mechanism, we consider the following two cases, where the miner chooses to cooperate or defect \cite{hu2017anti}.

a) if the miner is considered as cooperative, the pool may reward her, resulting in $E_m^{t+1}\ge E_m^t$. In this case, with the increase of $E_m^{t+1}$ and $f_m^{t+1}$, $W_c^{t+1}$ turns to
\begin{equation}\label{14}
  W_c^{t+1}=\frac{E_m^{t+1}-(1-f_m^{t+1})W_d^{t+1}}{f_m^{t+1}}.
\end{equation}

Hence, $\lim\limits_{t\to+\infty}W_c^{t+1}=\frac{E_m^{t+1}-(1-f_m^t)W_d^{t+1}}{f_m^t}>W_c^t$ because of $W_d^{t+1}=W_d^t$.

b) when the miner is regarded as a defective miner, then we have $E_m^{t+1}\le E_m^t$, and the decrease of $E_m^{t+1}$ and $f_m^{t+1}$ will lead to
\begin{equation}\label{15}
  W_d^{t+1}=\frac{E_m^{t+1}-f_m^{t+1}W_c^{t+1}}{1-f_m^{t+1}}.
\end{equation}

Comparably, $\lim\limits_{t\to+\infty}W_d^{t+1}= \frac{E_m^{t+1}-f_m^tW_c^{t+1}}{1-f_m^t} <W_d^t$ because of $W_c^{t+1}=W_c^t$.

To sum up, Case a) indicates that $W_c^t$ increases and $W_d^t$ remains unchanged and Case b) implies that $W_d^t$ declines while $W_c^t$ remains steady.
Thus, $\exists T^*\in \mathbb{Z^+}$, such that $\forall t>T^*$, $W_c^t>W_d^t$ holds. Based on this, $E_m^t$ can be derived as
\begin{equation}\label{16}
\begin{aligned}
E_m^t&= f_m^t W_c^t+(1-f_m^t) W_d^t \\
&< f_m^tW_c^t+(1-f_m^t) W_c^t=W_c^t.
\end{aligned}
\end{equation}

In light of \eqref{16}, we can conclude that $q^{t+1}=q^t\frac{W_c^t}{E_m^t}\to 1$ with the increase of game round $t$. That is to say, the memorial evolutionary miner will gradually increase the cooperation probability to one eventually.
\end{IEEEproof}

Conclusively, the non-memorial and memorial evolutionary miner will be encouraged to behave cooperatively by the proposed ZD incentive mechanism in the end.

Another essential nature of the proposed incentive mechanism is that it can be employed into the prepaid mechanism as well as the postpaid mechanism, with the former rewards the miner when a share is submitted and the latter defines the terminal of a mining round as the time a block is mined successfully. Noteworthily, the ZD incentive mechanism is free-fee charged for miners in both prepaid and postpaid cases due to their wholehearted devotions. More importantly, in the postpaid mechanism, the proposed incentive mechanism can hinder pool hopping attackers without putting any risk on the pools since our mechanism enables the miners to mine wholeheartedly until a block is generated successfully.

Now that such a powerful strategy the pool can employ, he has an overwhelmingly dominant position compared with the miner, then {\it is the pool capable of getting a higher payoff greedily through defecting when the miner collaborates?} We use the following theorem as a response to the above concern.

\begin{theorem}\label{5.3}
{\it When the miner chooses to cooperate, the only rational strategy of the pool who employs the ZD incentive mechanism is to collaborate.}
\end{theorem}
\begin{IEEEproof}
{As demonstrated in Theorems \ref{5.1} and \ref{5.2}, the miner will choose to contribute her maximum computational power into the pool because of the effectiveness of the proposed ZD incentive mechanism. In this case, the pool will provide the miner with the maximal payoff. Therefore, we will discuss what the ZD strategy is when the pool sets the expected payoff of the miner as the optimal value in the following.}

According to Section \ref{sec:sec4}, the miner's expected payoff can be set as $S_m=\frac{(1-p_1)S_m^{dd}+p_4S_m^{cc}}{1-p_1+p_4}$, which belongs to $[S_m^{dd},S_m^{cc}]$. Due to
\begin{equation}\label{piandao}
\begin{aligned}
  \frac{\partial S_m}{\partial p_1}&=\frac{p_4(\rho-\sigma)}{(1-p_1+p_4)^2},\\
  \frac{\partial S_m}{\partial p_4}&=\frac{(1-p_1)(\rho-\sigma)}{(1-p_1+p_4)^2},
\end{aligned}
\end{equation}
$\frac{\partial S_m}{\partial p_1}>0$ and $\frac{\partial S_m}{\partial p_4}>0$ because of $\rho>\sigma$ as indicated in Theorem \ref{2.2}, implying a monotonically increasing relationship between $S_m$ and $p_1$, $p_4$. Hence, when $p_1=1$, $p_4=1$, the pool can maximize the miner's expected payoff. Furthermore, according to \eqref{eq:pool_con_miner_p_23}, if $p_1$ and $p_4$ are equivalent to 1, the only possible value of $p_2$ is 1 because $p_2$ should lie in $[0,1]$ to be a probability, so as for $p_3$. That is to say, the pool can set $\mathbf p =(1,1,1,1)$ to maximize the payoff of a miner.

In light of the above analysis, once the miner cooperates, the pool will set his ZD strategy as $\mathbf p=(1,1,1,1)$ to maximize a collaborative miner's expected payoff. That is to say, whenever the miner cooperates, the pool will collaborate subsequently.

%As for a proportional ZD pool, he awards the miner due to the collaborative rounds continuously. If the miner cooperates more than $\nu$ rounds where $\nu\ge\frac{K_m}{k}$, the pool will offer the maximal expected payoff $K_m$ to the miner, whose strategy in this case is $\mathbf p =(1,1,1,1)$, which can be proved similar to the above case where an extreme ZD pool exists. Since the pool becomes a collaborative player until the miner has cooperated for $\nu$ rounds, the corresponding cooperative behavior for a proportional ZD pool is referred as delayed cooperation.
\end{IEEEproof}

In summary, the pool will be collaborative in return if the miner offers her maximum computing power. Thus, the proposed ZD incentive mechanism is fair to both sides, which makes it be long-term sustainable. Such an aim is achieved via controlling the miner's short-term expected payoff by the pool. Then, {\it what are the players' actual payoffs over the long run?} This question can be answered by the following two theorems.

\begin{theorem}\label{5.4}
{\it In the long run, the miner's actual payoff equals to $K_m$ based on our proposed ZD incentive mechanism.}
\end{theorem}
\begin{IEEEproof}
a) For a non-memorial evolutionary miner, $\exists \tau \in \mathbb{Z^+}$, such that $\forall t\ge\tau$, $q^t$ can be maximaized. That is to say, when $t\ge \tau$, the expected payoff of the miner is identical to $K_m$, which is the maximum payoff for a cooperative miner. In light of this, the actual payoff of the miner $P_m^A$ can be derived as the average of the expected payoff $E_m^i$ in each round $i$, where $i<\tau$ and the expected payoff $K_m$ after round $\tau$. Therefore, $P_m^A$ can be written as:

\begin{equation}\label{dingli3a}
  P_m^A=\lim\limits_{t\to\infty}\frac{\Sigma_{i=1}^{\tau-1}E_m^i+\Sigma_{i=\tau}^{t} K_m}{t}=K_m.
\end{equation}

b) The actual payoff of a memorial evolutionary miner is
\begin{equation}\label{dingli3b}
\begin{aligned}
  P_m^A&=\lim\limits_{t\to\infty}\frac{W_c^t-(1-p^t)(K_m+\sigma-\rho)}{p^t}
  =\lim \limits_{t\to\infty}W_c^t\\
  &=\lim\limits_{t\to\infty}\frac{E_m^t-(1-q^t)W_d^t}{q^t}=\lim \limits_{t\to\infty}E_m^t=K_m.
\end{aligned}
\end{equation}
\end{IEEEproof}

By inspecting Theorem \ref{5.4}, the miner will receive the actual payoff $P_m^A$ as $K_m$ over the long run. Then, {\it is it possible for the pool to own more payoff by greedy behavior?} This question can be resolved by the following theorem.

\begin{theorem}\label{5.5}
{\it In the long run, the pool's actual payoff $P_p^A$ is equivalent to $K_p$ based on our proposed ZD incentive mechanism.}
\end{theorem}
\begin{IEEEproof}
According to Theorem \ref{5.3}, the pool will behave cooperatively to reward a collaborative miner, implying that $XY=cc$ is the stable state for the game. In such a case, $P_p^A=K_p$ holds according to Table \ref{tb:payoff}.
\end{IEEEproof}

In light of Theorems \ref{5.4} and \ref{5.5}, the pool and miner will obtain the actual payoffs as $K_m$ and $K_p$, respectively. That is to say, neither the pool nor the miner can receive higher reward by noncooperative manner over the long run, which is quite fair for both sides.
\section{Performance Evaluation}\label{sec:evalu}
To testify the effectiveness of the ZD incentive mechanism proposed in Section \ref{sec:sec 5},  we conduct experimental simulations in this section.
To be specific, %we simulate the interplay between the pool and the miner in this subsection. The experiments conducted in the following are the cases where the pool is a ZD player while the miner is an arbitrary player. In the beginning,
we set the payoff vectors of the pool and the miner as $\mathbf S_p=(3,0,5,2)$ and $\mathbf S_m=(3,5,0,2),$ which is a typical example of the prisoner's dilemma. We also carry out the simulations with other parameter settings and derive the comparable results. So we omit to present those results to avoid redundancy. Note that each simulation is repeated 100 times to get the average value for statistical confidence.%{\color{red}the word is exactly the same as icdcs of huqin}

\begin{figure}[t]
\centerline{\includegraphics[width=2.5in]{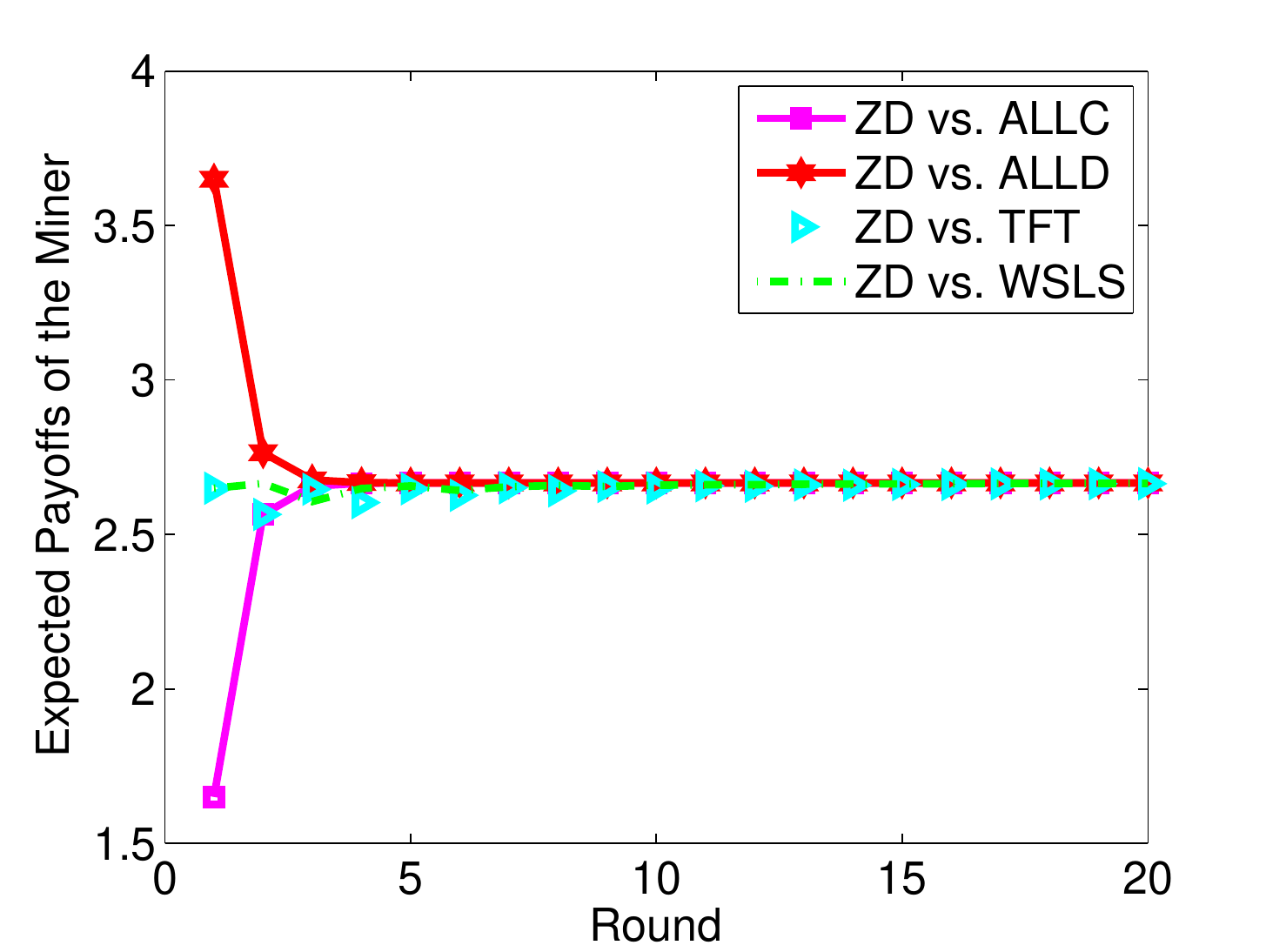}}
\caption{The expected payoffs of the miner when she adopts ALLC, ALLD, TFT, WSLS strategies and the pool employs the ZD strategy.}
\label{fig:1}
\end{figure}

In detail, if the pool is a ZD adopter competing with a miner who employs four classical strategies, i.e., ALLC,  ALLD, TFT and WSLS, the miner's expected payoffs can be set at a fixed value as shown in Fig. \ref{fig:1}. Taking the specific ZD strategy of the pool $\mathbf{p}=(0.9,0.3,0.8,0.2)$ as an example, no matter what strategies the miner employs, her expected payoff will finally become to a constant. That is to say, the adversary's outcome can be controlled unilaterally by the ZD adopter because of his effective strategy.

As mentioned in Section \ref{sec:sec 5}, the classical strategies ALLC, ALLD and TFT are out of our consideration because the strategies are irrelevant to the payoff of the player. Moreover, WSLS is regarded as a special evolutionary strategy. Hence, only the simulations of the evolutionary miners who compete with a ZD pool are included in this work, which are demonstrated as follows.

\begin{figure}[t]
\centering
\subfigure[$q^0=0.01$]{
\begin{minipage}[t]{0.45\linewidth}
\centering
\includegraphics[width=1.1\textwidth]{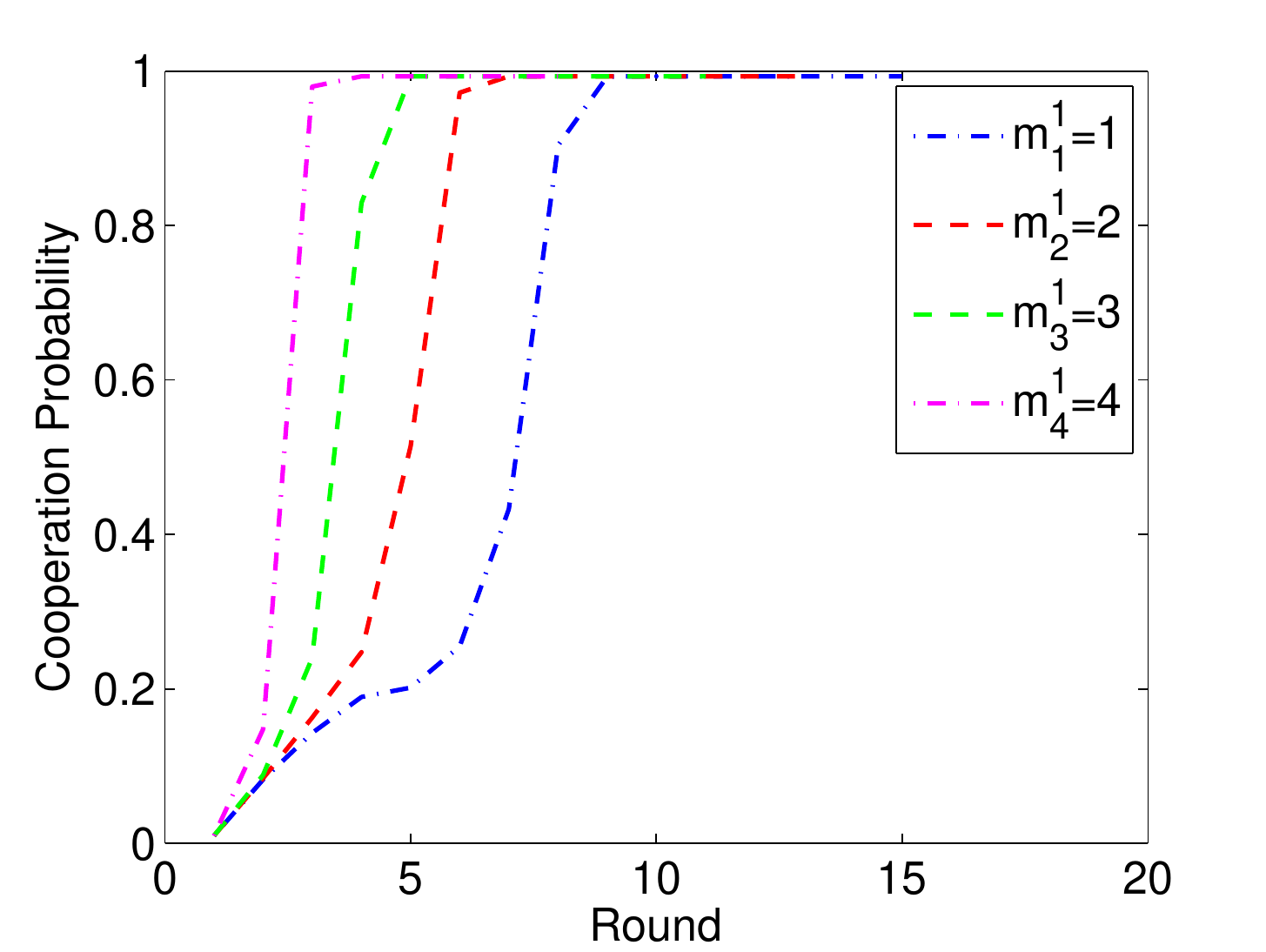}
\end{minipage}
}
\subfigure[$q^0=0.1$]{
\begin{minipage}[t]{0.45\linewidth}
\centering
\includegraphics[width=1.1\textwidth]{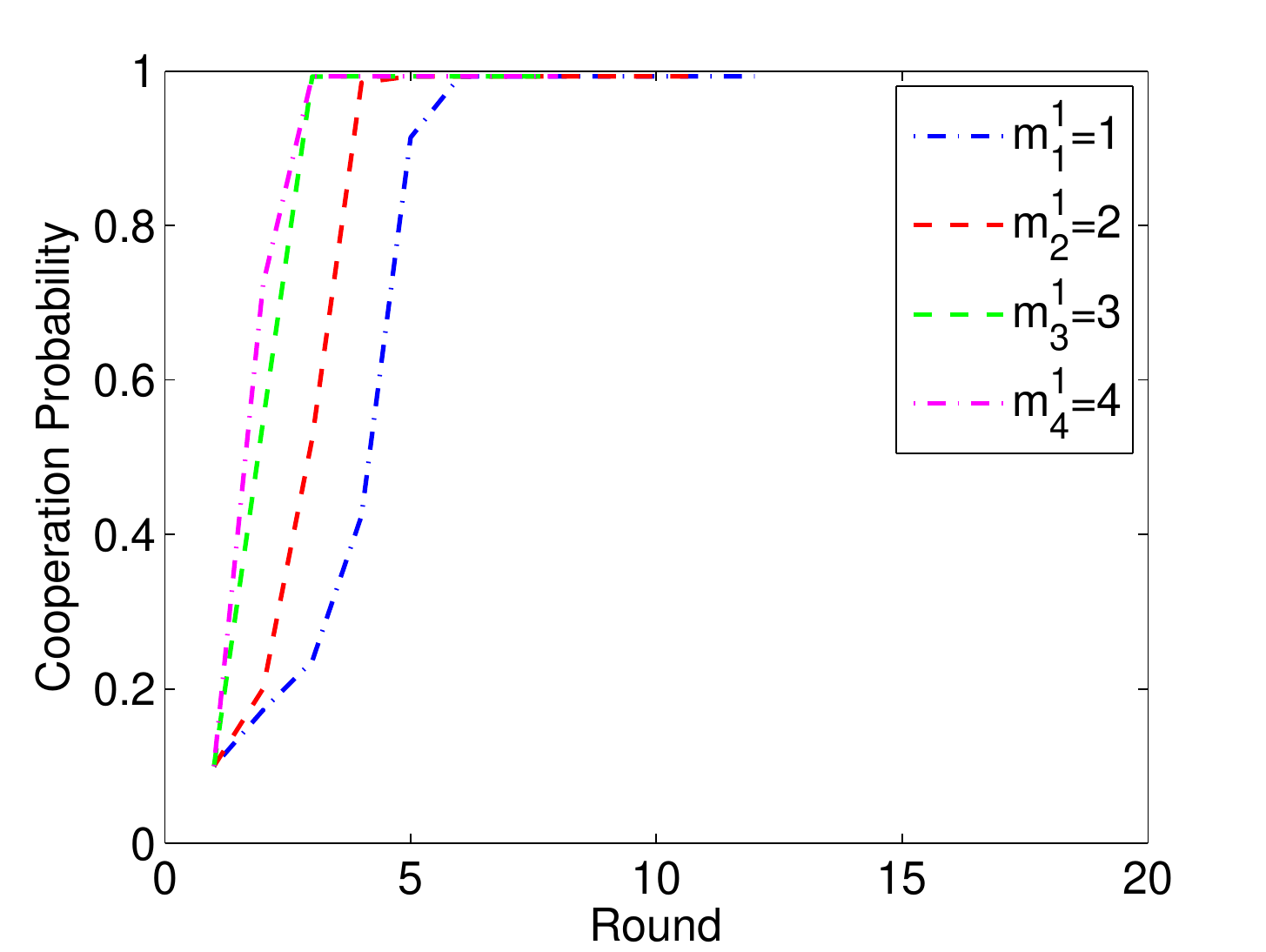}
\end{minipage}
}
\subfigure[ $q^0=0.5$]{
\begin{minipage}[t]{0.45\linewidth}
\centering
\includegraphics[width=1.1\textwidth]{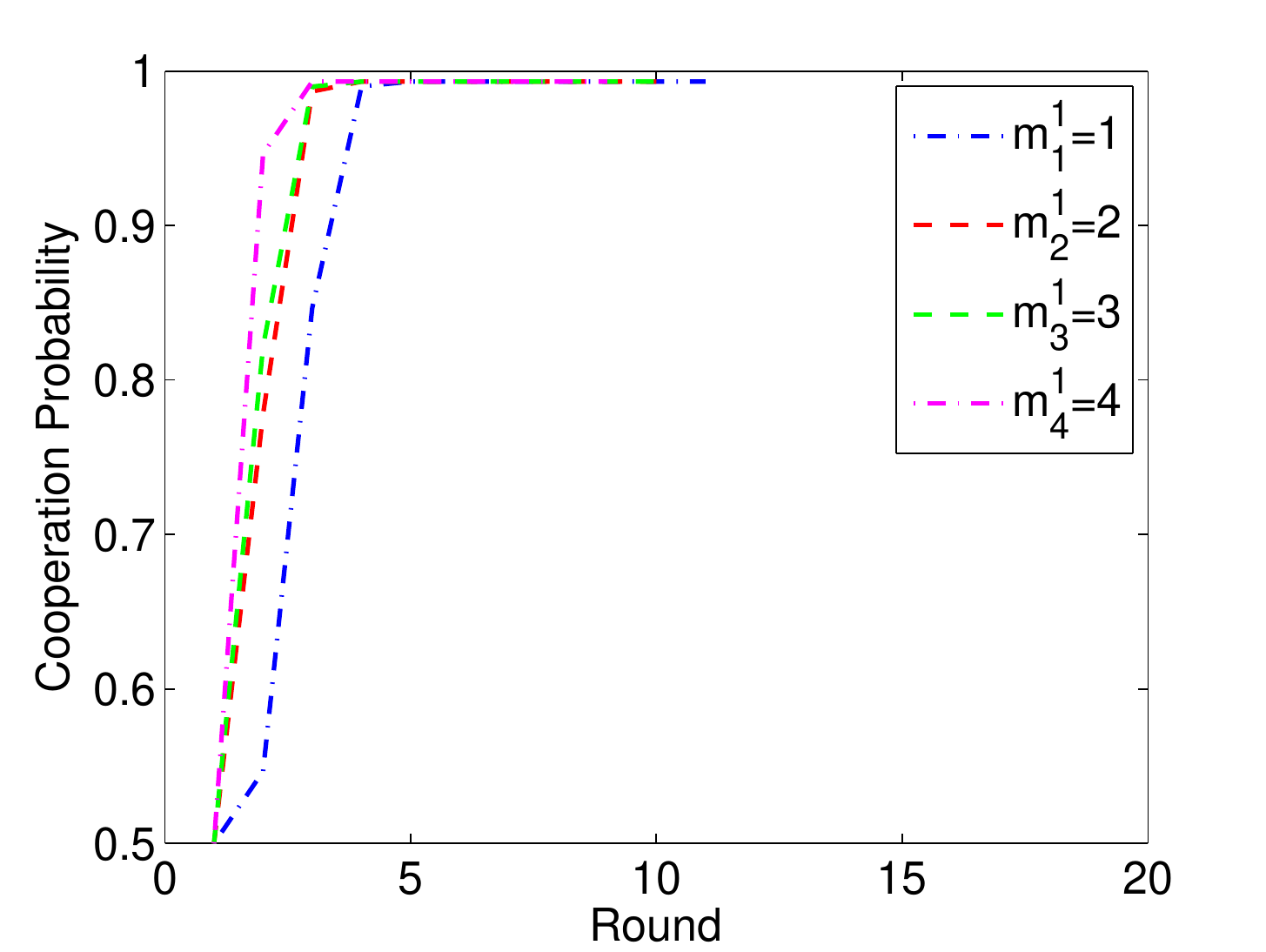}
\end{minipage}
}
\subfigure[$q^0=0.8$]{
\begin{minipage}[t]{0.45\linewidth}
\centering
\includegraphics[width=1.1\textwidth]{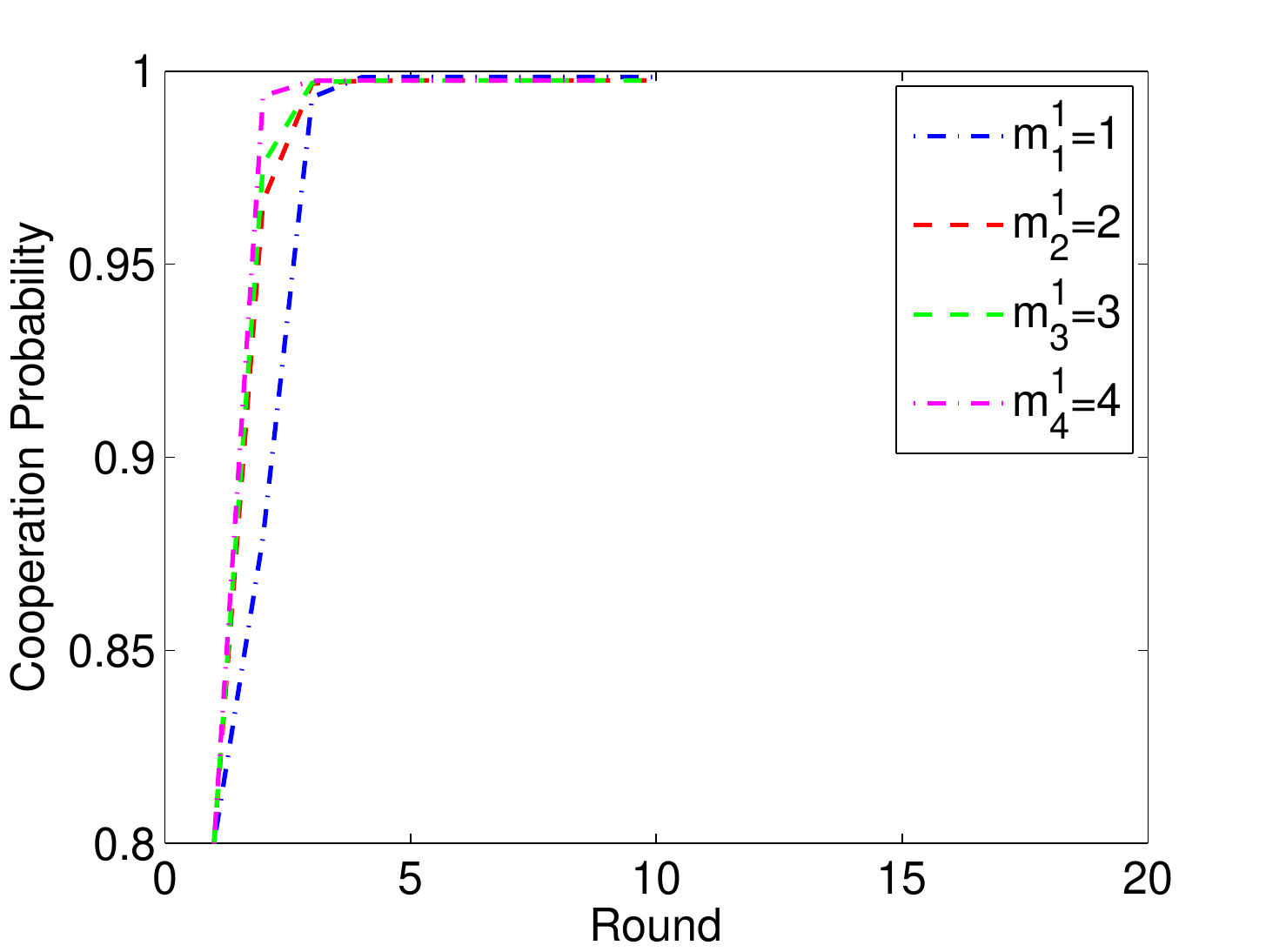}
\end{minipage}
}
\centering
\caption{The evolutions of the CPs of the non-memorial evolutionary miners  when $\epsilon=5$.}
\label{fig:2}
\end{figure}

\begin{figure}[t]
\centering
\subfigure[ $q^0=0.01$]{
\begin{minipage}[t]{0.45\linewidth}
\centering
\includegraphics[width=1.1\textwidth]{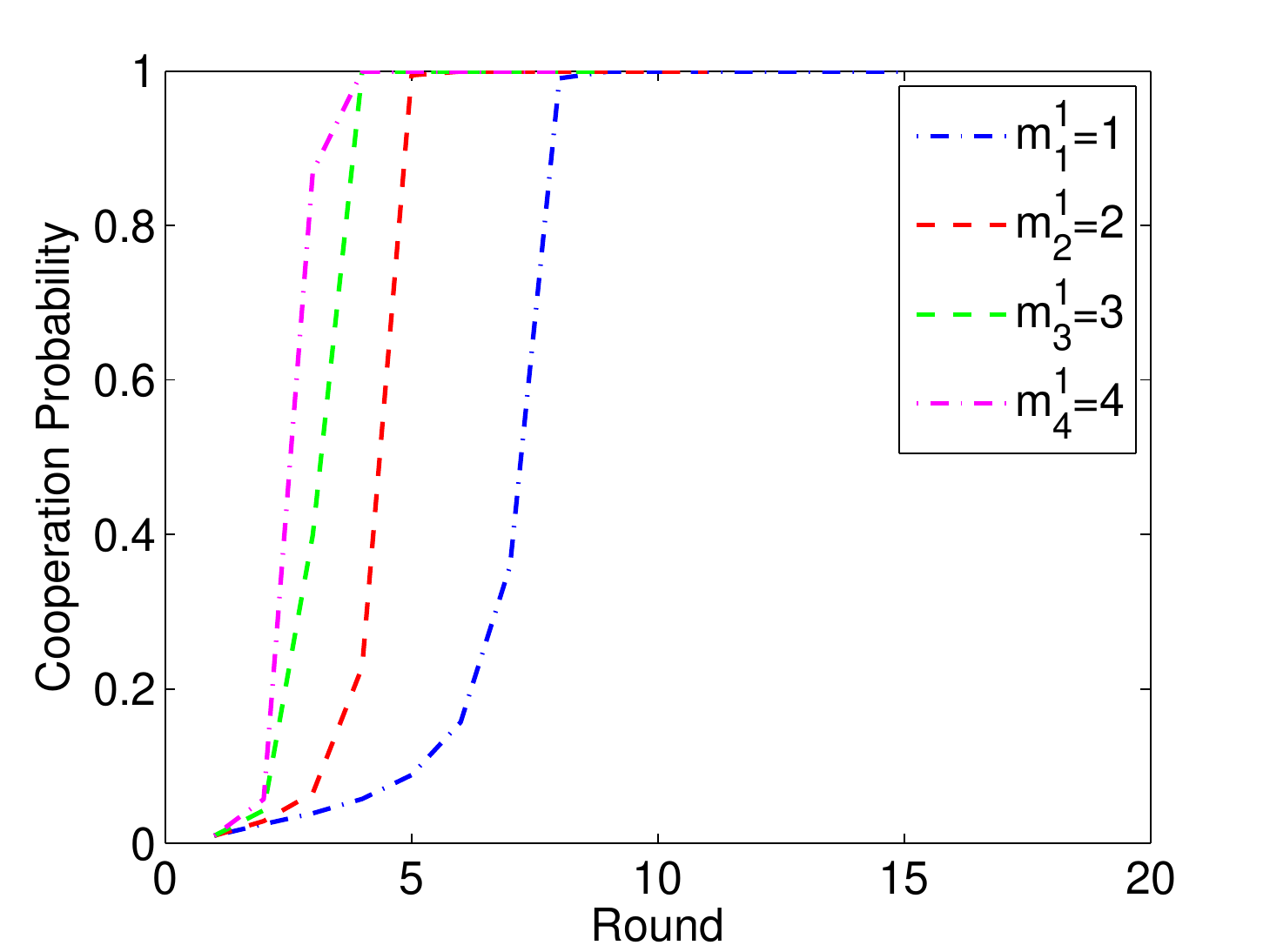}
\end{minipage}
}
\subfigure[ $q^0=0.1$]{
\begin{minipage}[t]{0.45\linewidth}
\centering
\includegraphics[width=1.1\textwidth]{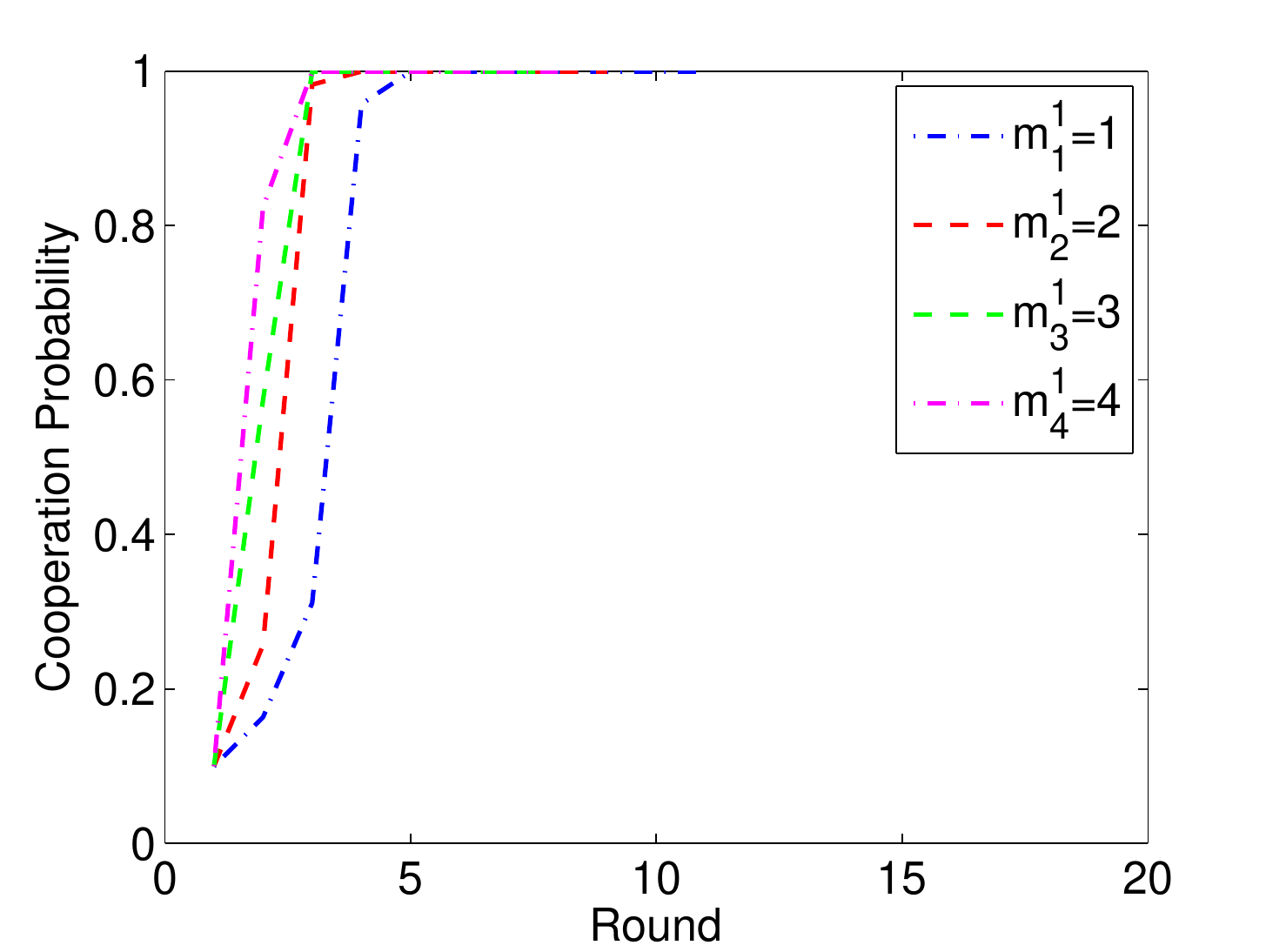}
\end{minipage}
}
\subfigure[$q^0=0.5$]{
\begin{minipage}[t]{0.45\linewidth}
\centering
\includegraphics[width=1.1\textwidth]{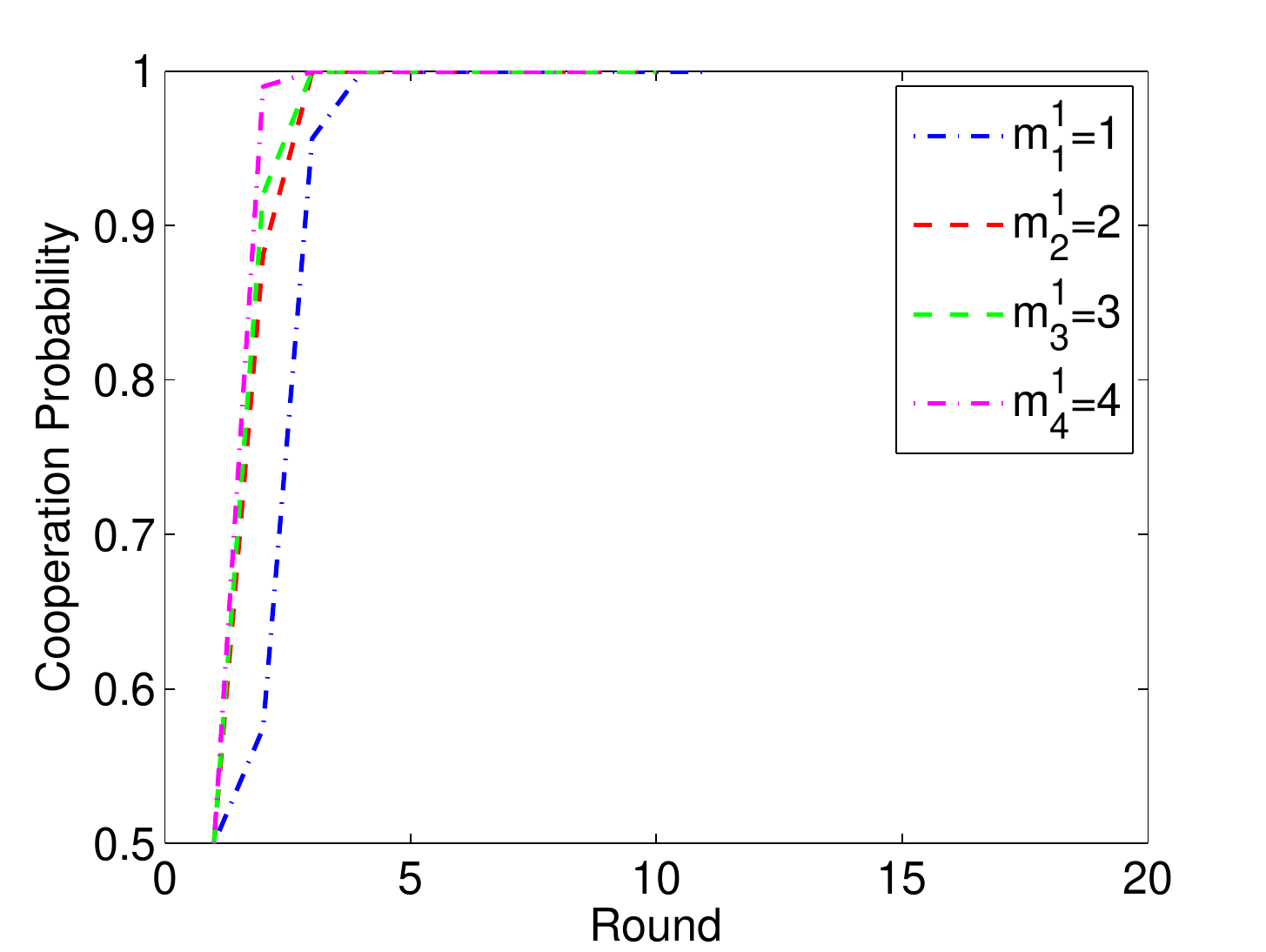}
\end{minipage}
}
\subfigure[$q^0=0.8$]{
\begin{minipage}[t]{0.45\linewidth}
\centering
\includegraphics[width=1.1\textwidth]{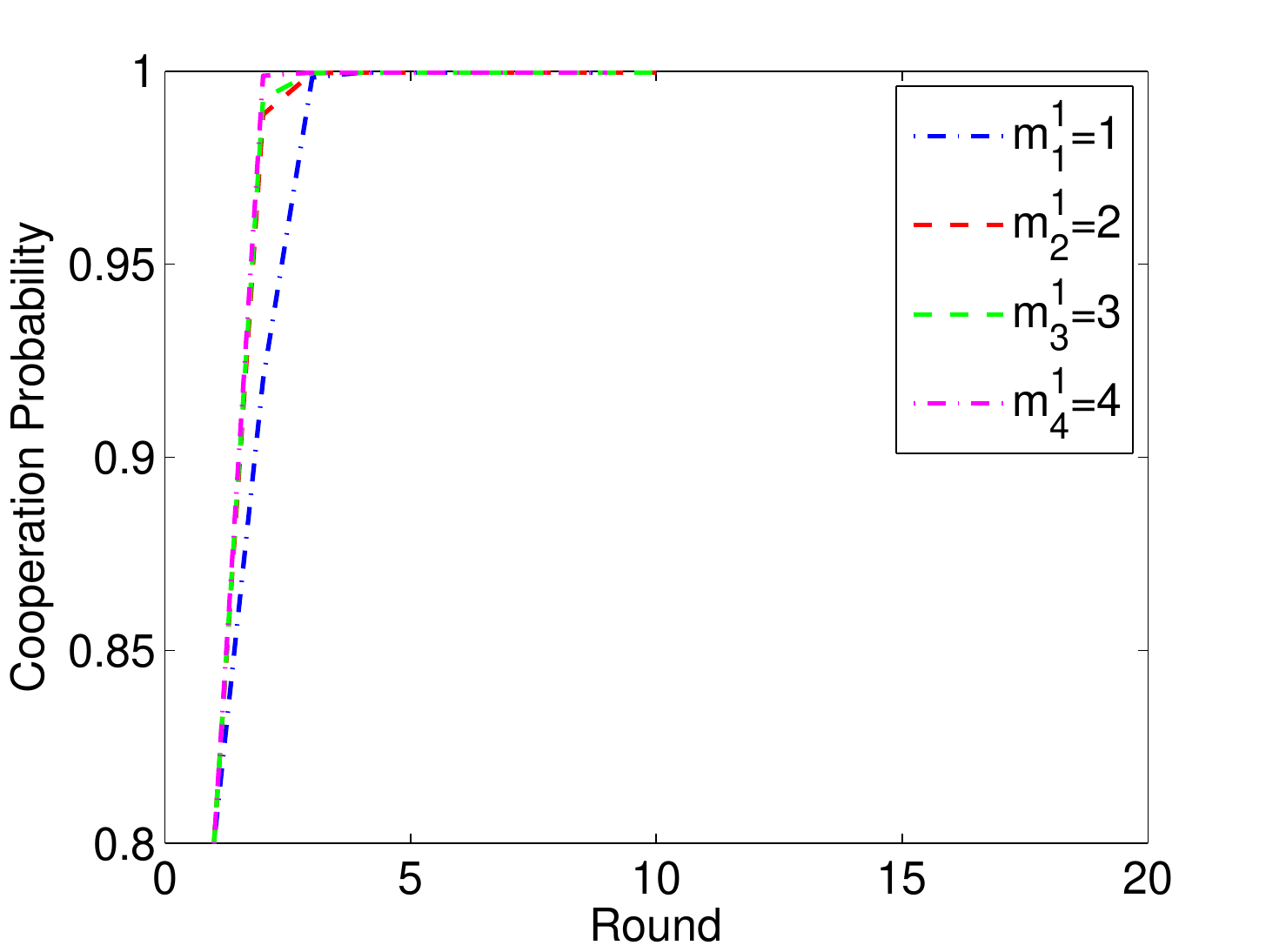}
\end{minipage}
}
\centering
\caption{The evolutions of the CPs of the non-memorial evolutionary miners  when $\epsilon=8$.}
\label{fig:3}
\end{figure}

In our simulation, we assume there are four miners  in a pool, whose initial computational powers are respectively  $m_1^1=1, m_2^1=2, m_3^1=3, m_4^1=4$ \footnote{The cases in which more miners exist in a ZD pool share the same conclusion, so we omit it for reducing repetition.}. Setting the original cooperation probabilities (CPs) $q^0=0.01, q^0=0.1, q^0=0.5$ and $q^0=0.8$, Figs. \ref{fig:2} and \ref{fig:3} respectively show how the CPs of the non-memorial evolutionary miners evolve according to  the proposed ZD incentive mechanism when $\epsilon$\footnote{ $\epsilon$ is set to be big enough here so that the maximum cooperation probability of a non-memorial evolutionary player (calculated by \eqref{eq:10}),  can approach to 1.}$=5$ and $8$.

Through further observation of Figs. \ref{fig:2} and \ref{fig:3}, we can conclude that the CPs of the non-memorial evolutionary miners converge to one with different speeds, which is mainly because of different initial computational investments and the scaling parameter $\epsilon$. To be specific, a miner with a larger initial computing investment would be more inclined to accelerate the cooperation process due to the higher growth of payoff. Intuitively, a higher $\epsilon$ brings about a faster convergence speed of the CP according to \eqref{eq:10}.

\begin{figure}[t]
\centering
\subfigure[$p^0=q^0=0.01$]{
\begin{minipage}[t]{0.45\linewidth}
\centering
\includegraphics[width=1.1\textwidth]{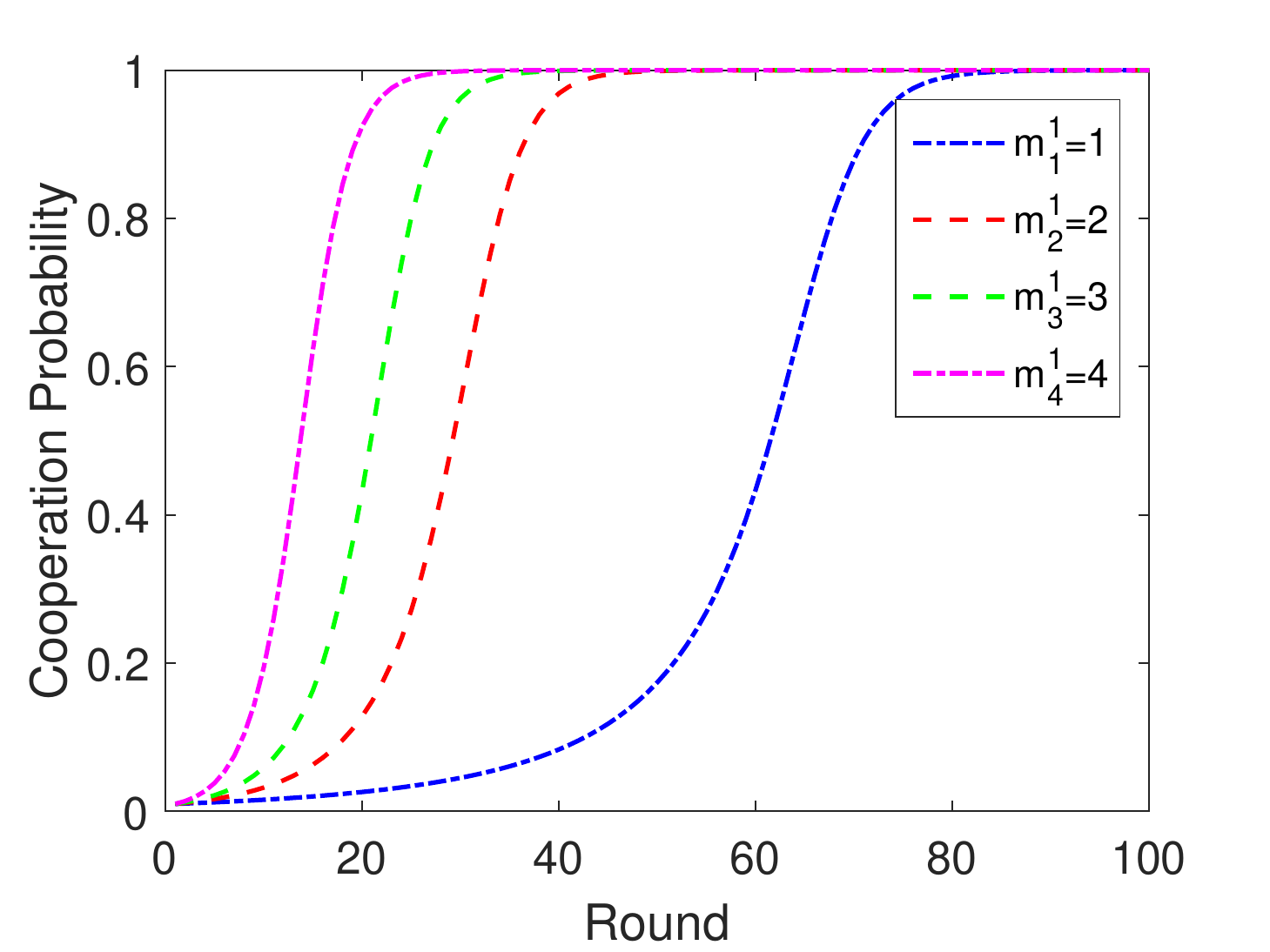}
\end{minipage}
}
\subfigure[ $p^0=q^0=0.1$]{
\begin{minipage}[t]{0.45\linewidth}
\centering
\includegraphics[width=1.1\textwidth]{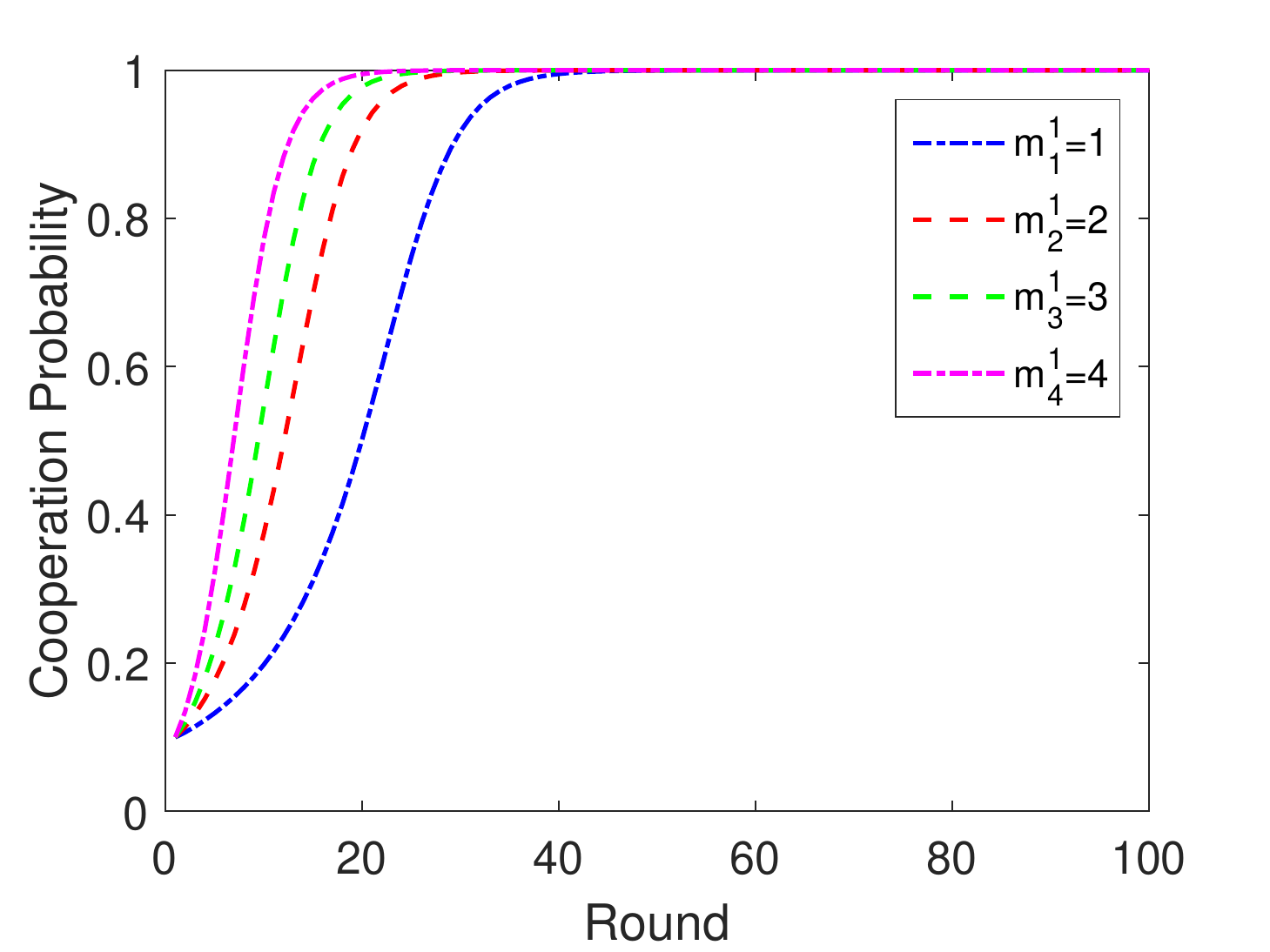}
\end{minipage}
}
\subfigure[$p^0=q^0=0.5$]{
\begin{minipage}[t]{0.45\linewidth}
\centering
\includegraphics[width=1.1\textwidth]{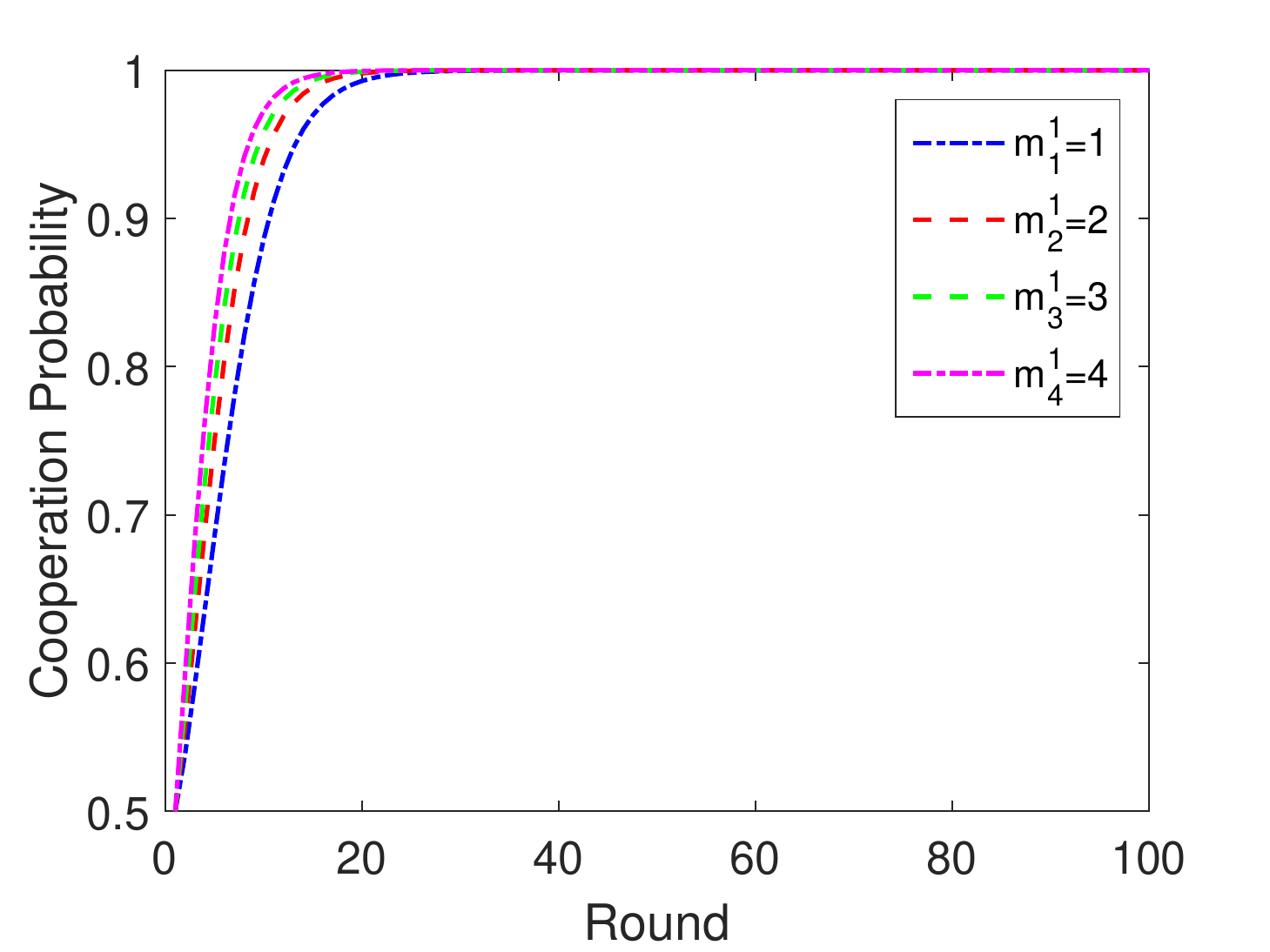}
\end{minipage}
}
\subfigure[$p^0=q^0=0.8$]{
\begin{minipage}[t]{0.45\linewidth}
\centering
\includegraphics[width=1.1\textwidth]{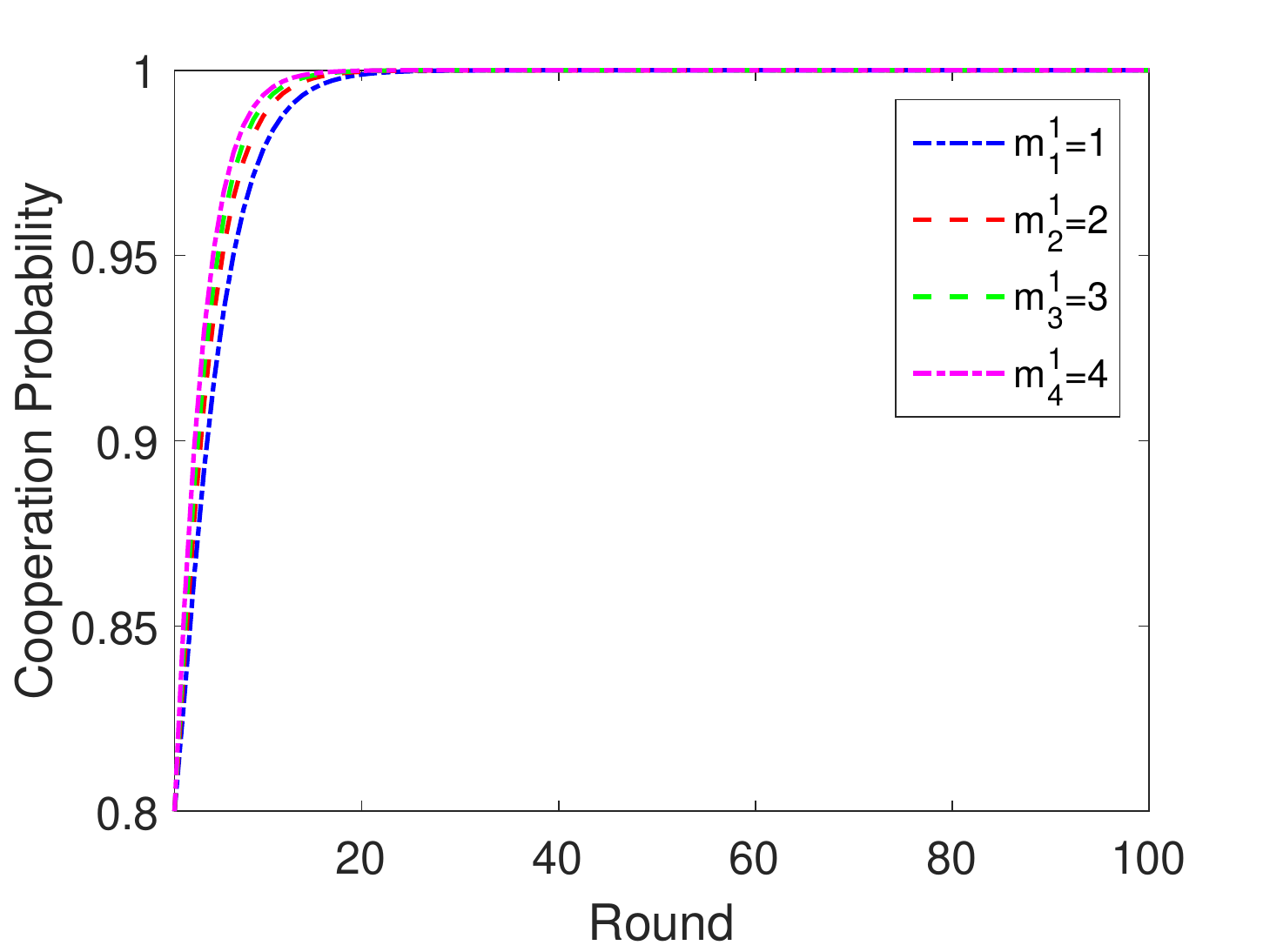}
\end{minipage}
}
\centering
\caption{The evolutions of the CPs of the memorial evolutionary miners.}
\label{fig:4}
\end{figure}

Fig. \ref{fig:4}  plots the CPs of a memorial evolutionary miner driven by the ZD-based incentive mechanism, where the CPs go up to 1 gradually with the initial values as $p^0=q^0=0.01, 0.1, 0.5,$ and $0.8$. In detail, each subfigure shows that the CP of the miner with a small initial input converges slowly compared with other miners, even though they share the same initial cooperation probability. The reason may lie in that the miner with a smaller initial computing investment may get a relatively lower payoff in the beginning, leading a slow growth of the expected payoff. Thus, her CP would rise slower comparably. Moreover, considering the CPs of a miner with the same initial investment but having different initial cooperation probabilities, for example, the blue lines in subfigures (a)-(d), the result is that the higher the initial CP is, the faster it is converged to one, which is mainly caused by the {\it memory} we mentioned above in light of \eqref{eq:memo}.

%In summary, both the theoretical analysis and the simulation results demonstrate the effectiveness of the proposed incentive mechanisms based on the ZD theory. Thus, equipped with the dominant incentive mechanisms, the pool is capable of coercing the miner to behave cooperatively and hindering the pool-hopping attack successfully. While, in this case, the pool's strategy is to collaborate, leading $XY=cc$ as the stable state at last. As a consequence, the incentive mechanisms proposed in our work enable mutual cooperation for both players, making that (1) the miner can benefit from not being charged fees by the pool since she will behave cooperatively, and (2) the pool can get higher revenue because of the miner's wholehearted contributions.

%Noteworthily, two incentive mechanisms based on the ZD strategy proposed in this paper might be selectively employed for different types of players. To be specific, the proportional ZD incentive mechanism provides a smooth reward mechanism, which is designed for the miners who desire a steady income with a lower variance. On the contrary, for the players who have intense demands for mutual cooperation, the extreme ZD incentive mechanism is suitable for them since it can speed up the process of collaboration to a great extent.
\section{Related Work}\label{sec:sec2}
At present, the researchers mainly focus on three kinds of security attacks in pooled mining: the selfish mining attack, the block withholding (BWH) attack and the pool-hopping attack.

In detail, a selfish mining attacker\cite{majority} keeps its mined block secret and intentionally forks the main blockchain. Specifically, the selfish miner mines on its private branch instead of working on the public chain as the honest miners. When the public ledger approaches it's private chain, the selfish miner advertises its concealed chain to the public, leading to wasting resources of the honest miners on resolving cryptopuzzles which ends up gaining no rewards. Several defense mechanisms have been proposed to block this selfish manner as well as its variants. For example, Saad {\it et al.}\cite{selfish1} developed a defense mechanism in the network-wide scope to detect and deter selfish miners; Zhang {\it et al.} \cite{selfish2} proposed a backward-compatible mechanism to defend selfish attacks.

The BWH attackers pretend to devote their computational capabilities into the target pool and then obtain payoffs. However, they send only partial proof of work, not full proof of work, resulting in reward reduction to other miners in the pool. This kind of attack was first proposed in \cite{analysis}, after which, Courtois {\it et al.}\cite{courtois} summarized its concept and Eyal modeled the confrontation between the pools as a prisoner's dilemma in \cite{miner}. Specifically, in \cite{miner}, a Nash equilibrium was established, where the rational pools would attack each other, resulting in a lose-lose situation. Besides, the pools are trapped into an iterative prisoner's dilemma, in which the pool chooses to attack or not is the so called miner's dilemma. Ongoing researches on avoiding this attack have proposed some efficient and cheap defense mechanisms. For example, Bag {\it et al.} in\cite{atrans} proposed a generic scheme to counter BWH attacks via employing cryptographic commitment schemes, based on which, an implementation using hash function was presented as an alternative. Besides, Luu {\it et al.}\cite{splitting} put forward a power splitting game for the miners so as to find a solution to fight back the BWH attacks. Additionally, Hu {\it et al.} \cite{hu2019} took advantages of the Zero-determinant theory to analyze the BHW attacks between two pools. Based on which, different conditions for the pools playing the ZD strategy individually and simultaneously have been demonstrated comprehensively.

We focus on the pool-hopping attack in this work. Pioneer countermeasures are PPS, PPLNS and their variants, including the Slush's method, maximum pay-per-share (MPPS), and pay-once-PPLNS. Detailedly, the pool manager can calculate the score of each share based on the exponential score function $s=e^{\frac{T}{c}}$, in which $s$ represents the score of the share given in time $T$ and $c$ denotes the scaling parameter. Due to the share's score, the pool hopping behavior can be alleviated in mining pools by reducing the score of shares at the earlier stage of the round while increasing the score of shares later on. Such kind of score-based method is recognized as the Slush's method and has been applied in the mining pools such as Slushpool\cite{slushpool}. Besides, in the maximum pay-per-share method, two balances are kept for each miner, that is, a PPS balance and a proportional balance\cite{analysis}. To be specific, if the miner offers a share, her PPS share balance is increased as if the pool is a PPS pool. When the pool generates a block, the proportional balances of  the miners are increased as if they have joined a proportional pool. Based on which, the reward paid for each miner is the minimum between the PPS balance and the proportional balance. In pay-once-PPLNS, every share is rewarded at most once\cite{analysis}. In other words, the share is deleted after it is paid, leading a higher probability to the elder shares to be paid for future blocks. If a share is partially paid, it will be deleted partially. %However, all the above variants are capable of alleviating the pool-hopping attack, but not hopping-proof completely.
However, theoretical analysis on the above mechanisms are lacking and their effectiveness in preventing pool-hopping attacks still remain an open issue\cite{survey}.

\section{Conclusion}\label{sec:sec_conclude}
In this paper, we propose a hopping-proof pooled mining with fee-free in Blockchain. To that aim, we formulate the interaction between the pool and any miner as an IPD game and identify the corresponding conditions. The generality of our model capacitates the proposed pooled mining to have wide  applicability. Based on the model, we take advantage of the ZD theory to empower  the pool can unilaterally control the miner's payoff, which can be used to motivate the cooperation of  the non-memorial and memorial evolutionary miners through the proposed ZD incentive mechanism. Both theoretical and experimental analyses demonstrate the effectiveness of the ZD incentive mechanism. To the best of our knowledge, we are the first to propose a hopping-proof pooled mining with the  natures of {\it fee-free, wide applicability and fairness} at the same time.

% use section* for acknowledgment
%\ifCLASSOPTIONcompsoc
  % The Computer Society usually uses the plural form
  %\section*{Acknowledgments}
%\else
  % regular IEEE prefers the singular form
  %\section*{Acknowledgment}
%\fi
% Can use something like this to put references on a page
% by themselves when using endfloat and the captionsoff option.
\ifCLASSOPTIONcaptionsoff
  \newpage
\fi

\bibliographystyle{IEEEtran}
\bibliography{reference}

\begin{IEEEbiography}[{\includegraphics[width=1in,height=1.25in,clip,keepaspectratio]{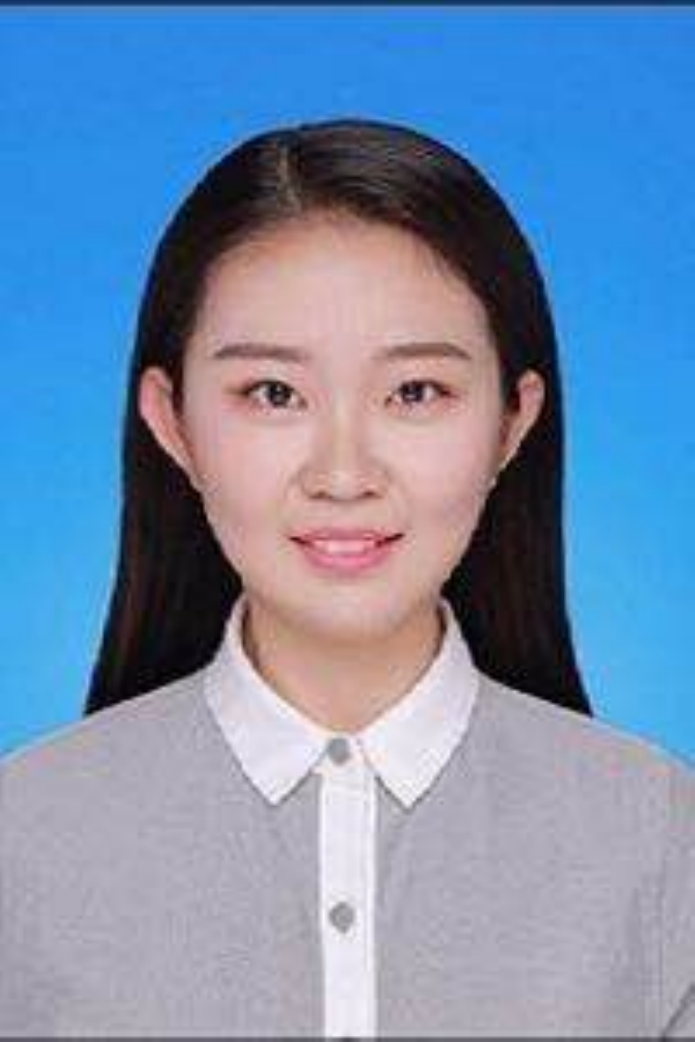}}]
{Hongwei Shi} received her B.S. degree in Computer Science from Beijing Normal University in 2018. Now she is pursuing her M.S. degree in Computer Science from Beijing Normal University. Her research interests include blockchain,
game theory and combinatorial optimization.
\end{IEEEbiography}

\begin{IEEEbiography}[{\includegraphics[width=1in,height=1.25in,clip,keepaspectratio]{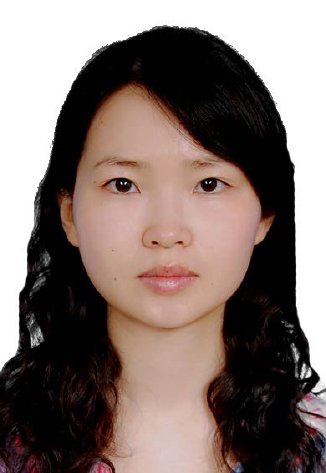}}]
{Shengling Wang} is a full professor in the School of Artificial Intelligence, Beijing Normal University. She received her Ph.D. in 2008 from Xi’an Jiaotong University. After that, she did her postdoctoral research in the Department of Computer Science and Technology, Tsinghua University. Then she worked as an assistant and associate
professor from 2010 to 2013 in the Institute of Computing Technology of the Chinese Academy of Sciences. Her research interests include mobile/wireless networks, game theory, crowdsourcing.
\end{IEEEbiography}

\begin{IEEEbiography}[{\includegraphics[width=1in,height=1.25in,clip,keepaspectratio]{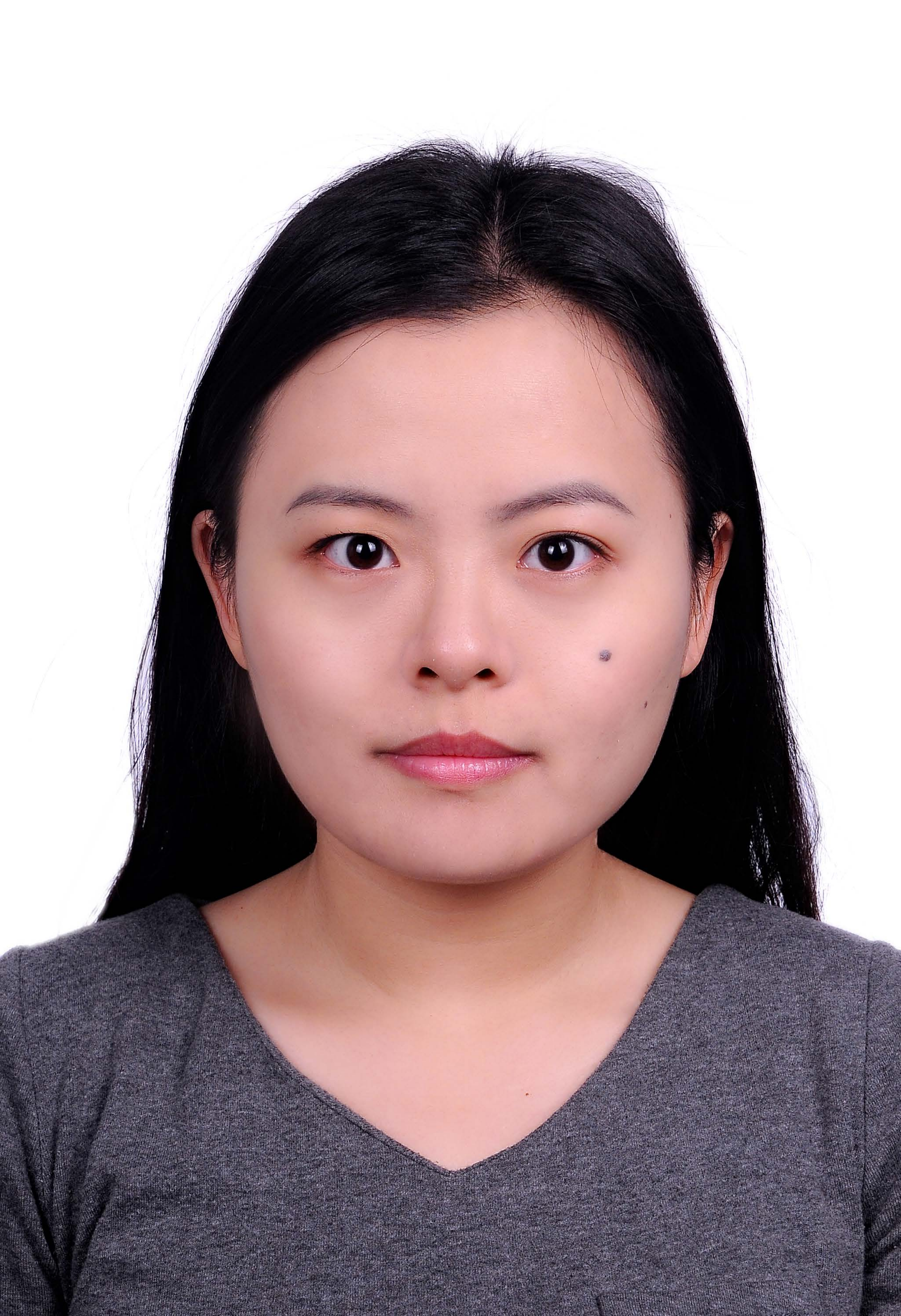}}]
{Qin Hu} received her Ph.D. degree in Computer Science from the George Washington University in 2019. She is currently an Assistant Professor in the department of Computer and Information Science, Indiana University - Purdue University Indianapolis. Her research interests include wireless and mobile security, crowdsourcing/crowdsensing and blockchain.
\end{IEEEbiography}

\begin{IEEEbiography}[{\includegraphics[width=1in,height=1.25in,clip,keepaspectratio]{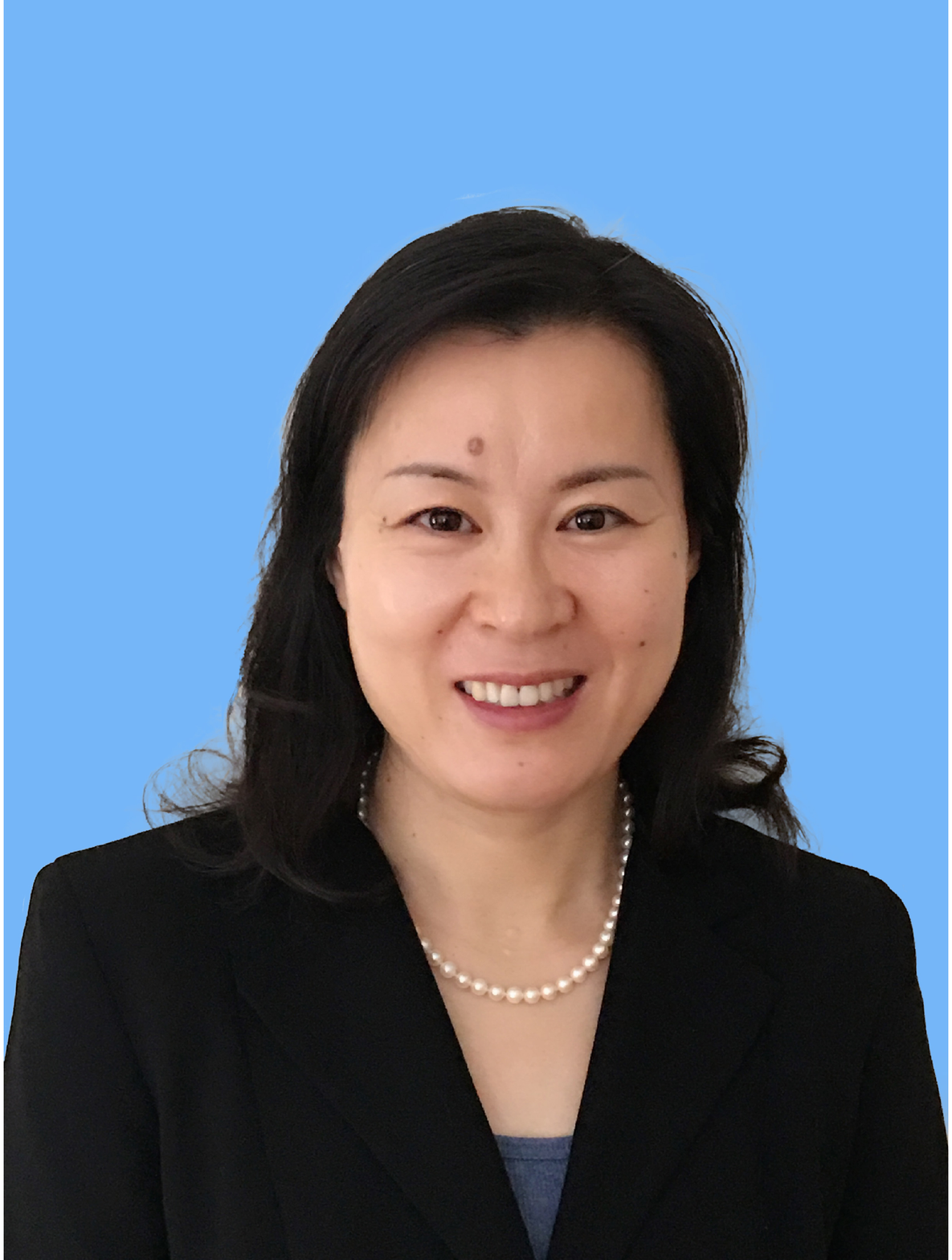}}]
{Xiuzhen Cheng}[F] received her M.S. and Ph.D. degrees in computer science from the University of Minnesota Twin Cities in 2000 and 2002. She is a professor in the Department of Computer Science, George Washington University, Washington, DC. Her current research interests focus on privacy-aware computing, wireless and mobile security, dynamic spectrum access, mobile handset networking systems (mobile health and safety), cognitive radio networks, and algorithm design and analysis. She has served on the Editorial Boards of several technical publications and the Technical Program Committees of various professional conferences/workshops. She has also chaired several international conferences. She worked as a program director for the U.S. National Science Foundation (NSF) from April to October 2006 (full time), and from April 2008 to May 2010 (part time). She published more than 170 peer-reviewed papers.
\end{IEEEbiography}

\begin{IEEEbiography}[{\includegraphics[width=1in,height=1.25in,clip,keepaspectratio]{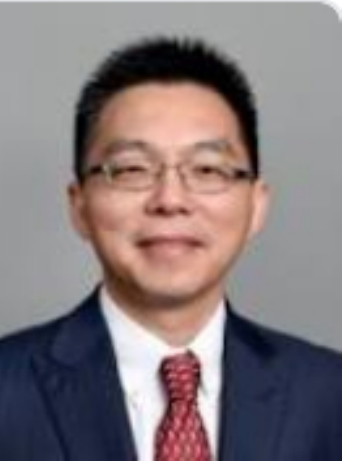}}]
{Junshan Zhang} received the Ph.D. degree from the School of Electrical and Computer Engineering, Purdue University, West Lafayette, IN, USA, in 2000. In August 2000, he joined the School of Electrical, Computer and Energy Engineering, Arizona State University, Tempe, AZ, USA, where he has been a Professor since 2010. His research interests fall in the general field of information networks and its intersections with power networks and social networks. His current research focuses on fundamental problems in information networks and energy networks, including modeling and optimization for smart grids, optimization/control of mobile social networks and cognitive radio networks, and privacy/security in information networks.
\end{IEEEbiography}

\begin{IEEEbiography}[{\includegraphics[width=1in,height=1.25in,clip,keepaspectratio]{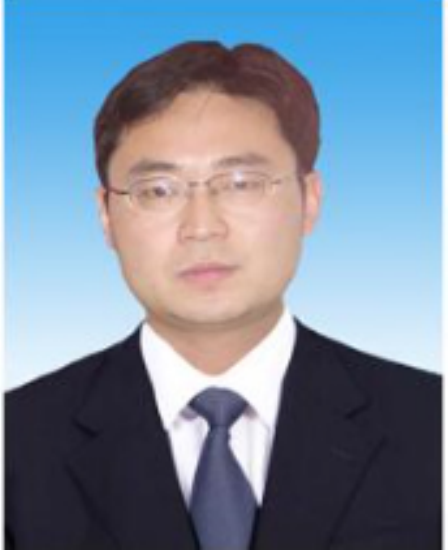}}]
{Jiguo Yu} received the Ph.D. degree from Shandong University, in 2004. He became a Full Professor with the School of Computer Science, Qufu Normal University, Shandong, China, in 2007. He is currently a Full Professor with the Qilu University of Technology (Shandong Academy of Sciences), Shandong Computer Science Center (National Supercomputer Center in Jinan), and a Professor with the School of Information Science and Engineering, Qufu Normal University. His research interests include privacy-aware computing, wireless networking, distributed algorithms, peer-to-peer computing, and graph theory. He is a member of ACM and a Senior Member of CCF.
\end{IEEEbiography}
\end{document}